\documentclass[a4paper,11pt]{article}
\usepackage{jheppub} 

\usepackage[T1]{fontenc}
\usepackage{epsfig}
\usepackage{graphicx}
\usepackage{hyphenat}
\usepackage{amsmath}
\usepackage{amssymb}
\usepackage{bbold}
\usepackage{mathtools}
\usepackage{mathrsfs}
\usepackage{slashed}
\usepackage{epstopdf}
\usepackage{xcolor}
\usepackage{braket}
\usepackage{subfigure}
\definecolor{lcolor}{rgb}{0.,0.0,0.}
\definecolor{citcolor}{rgb}{0,0.,0.5}

\usepackage[e]{esvect}
\usepackage{media9}

\newcommand{\secn}[1]{Section~1}
\newcommand{\appn}[1]{Appendix~1}

\long\def\comment#1{ }

\def\Tr{\text{Tr}}

\def\and{\quad\text{and}\quad}

\def\0{{\boldsymbol 0}}
\def\1{{\boldsymbol 1}}

\def\0{{\boldsymbol 0}}

\renewcommand\a{\alpha}

\renewcommand{\part}{{\rm part}}

\newcommand{\be}{\begin{equation}}
\newcommand{\ee}{\end{equation}}
\newcommand{\bes}{\begin{subequations}}
\newcommand{\ees}{\end{subequations}}
\newcommand{\bea}{\begin{eqnarray}}
\newcommand{\eea}{\end{eqnarray}}

\newcommand{\nn}{\nonumber \\}

\def\bea#1\eea{\begin{align}#1\end{align}}
\newcommand{\bef}{\begin{figure}[h!tb]\centering}
\newcommand{\eef}{\end{figure}}

\allowdisplaybreaks

\title{Hadronic scattering in (1+1)D SU(2) lattice gauge theory from tensor networks}

\author[a]{Jo\~{a}o Barata,}
\emailAdd{joao.lourenco.henriques.barata@cern.ch}
\affiliation[a]{CERN, Theoretical Physics Department, CH-1211, Geneva 23, Switzerland}

\author[b]{Juan Hormaza,}
\emailAdd{jhormaza@unal.edu.co}
\affiliation[b]{Universidad Nacional de Colombia, Manizales, Colombia.}

\author[c,d,e]{Zhong-Bo Kang,}
\emailAdd{zkang@physics.ucla.edu}
\affiliation[c]{Department of Physics and Astronomy, University of California,
Los Angeles, CA 90095, USA}
\affiliation[d]{Mani L. Bhaumik Institute for Theoretical Physics, University of California,
Los Angeles, CA 90095, USA}
\affiliation[e]{Center for Quantum Science and Engineering, University of California, Los Angeles, CA 90095, USA}

\author[f,g]{and Wenyang Qian}
\emailAdd{wqian@ccnu.edu.cn}
\affiliation[f]{Institute of Particle Physics and Key Laboratory of Quark and Lepton Physics (MOE),
Central China Normal University, Wuhan, 430079, Hubei, China}
\affiliation[g]{Instituto Galego de F\'isica de Altas Enerx\'ias IGFAE, Universidade de Santiago de Compostela,
E-15782 Galicia-Spain}

\preprint{CERN-TH-2025-214}

\abstract{
We present a first real-time study of hadronic scattering in a (1+1)-dimensional SU(2) lattice gauge theory with fundamental fermions using tensor-network techniques. 
Working in the gaugeless Hamiltonian formulation---where the gauge field is exactly integrated out and no truncation of the electric flux is required---we investigate scattering processes across sectors of fixed global baryon number $B = 0, 1, 2$. 
These correspond respectively to meson--meson, meson--baryon, and baryon--baryon collisions. 
At strong coupling, the $B = 0$ and $B = 2$ channels exhibit predominantly elastic dynamics closely resembling those of the U(1) Schwinger model. 
In contrast, the mixed $B = 1$ sector shows qualitatively new behavior: meson and baryon wavepackets become entangled during the collision, and depending on their initial kinematics, the slower state becomes spatially delocalized while the faster one propagates ballistically. 
We characterize these processes through local observables, entanglement entropy, and the information-lattice, which together reveal how correlations build up and relax during the interaction. 
Our results establish a first benchmark for non-Abelian real-time scattering from first principles and open the path toward quantum-simulation studies of baryon-number dynamics and inelastic processes in gauge theories.
}

\begin{document}

\maketitle

\section{Introduction}\label{sec:introduction}
A large part of our understanding of the matter content and interactions in the Standard
Model (SM) has been driven by high-energy scattering experiments over the last decades at several facilities worldwide~\cite{P5:2023wyd,NSAC2023_LRP}. This program has not only led to a better description of the fundamental properties of the theory but also to new insights into the many-body aspects of quantum field theories, essential for describing matter under extreme conditions, such as those expected to have existed in the early Universe. Nonetheless, the real-time, non-perturbative nature of scattering processes in gauge theories remains poorly understood. These aspects cannot be fully addressed within traditional approaches such as perturbation theory or Euclidean lattice field theory, due to the lack of convergence of perturbative series and the presence of sign problems, respectively; see e.g.~\cite{Asakawa:2000tr,Bruno:2020kyl}. 
Understanding real-time dynamics in gauge theories therefore represents one of the central open challenges in quantum field theory.

As a result, new theoretical and computational methods that enable the study of
real-time properties of scattering processes from first principles within quantum field
theory can play a decisive role in advancing our understanding of dynamical processes in gauge theories.
In the past decade, developments in quantum information science (QIS) have opened
powerful new avenues toward this goal; see
e.g.~\cite{Zohar:2021nyc,Banuls:2019bmf,Funcke:2023jbq,Halimeh:2025vvp}.
These approaches---including quantum computers, analog quantum simulators, and tensor
network methods---allow one to emulate the quantum properties of gauge theories using
either programmable quantum devices or classical (variational) algorithms. 
Among them, tensor-network formulations have proven particularly effective for
(1+1)-dimensional systems~\cite{Banuls:2018jag}, where they provide high-precision access to the
non-equilibrium dynamics of strongly coupled gauge theories in real-time.

In the context of particle scattering, recent efforts have achieved the first simulations
of elastic and inelastic processes in Abelian gauge theories using tensor-network and
quantum-simulation techniques, 
see e.g.~\cite{Papaefstathiou:2024zsu,Chai:2023qpq,Bennewitz:2025nhz,
Davoudi:2025rdv,Farrell:2025nkx,Barata:2025hgx,Abel:2025zxb,Carena:2024peb,
Jha:2024jan,Zemlevskiy:2024vxt,Ingoldby:2025bdb,Schuhmacher:2025ehh,Su:2024uuc,Belyansky:2023rgh}, 
and also~\cite{Pavesic:2025nwm,Cobos:2025krn,DiMarcantonio:2025cmf,
Gonzalez-Cuadra:2024xul,Joshi:2025rha,Xu:2025abo} for related extensions to (2+1)D models. 
However, so far the focus has been mainly restricted to Abelian theories. 
Extending this program to non-Abelian models represents a crucial step toward
capturing phenomena such as confinement and baryon formation, that
are central to QCD-like dynamics.

In this work we expand on these recent results and consider scattering processes in the (non-Abelian) SU(2) lattice gauge theory in (1+1)D. The most interesting difference compared to the Abelian models, is the existence of an extra conserved quantum number --- the baryon number $B$ --- which allows, for example, for the formation of (di-quark) baryon states absent in U(1). More generally, sectors of fixed $B$ admit more complex matter states, while in U(1) the particle spectrum is controlled by the lowest meson state (the Schwinger boson). The main goal of this paper is thus to explore the properties of scattering events at different $B$ values. In particular, we consider scattering processes for global $B=0,1,2$, i.e. meson-meson, baryon-meson, and baryon-baryon scattering, respectively. As we show below, and working close to the strong coupling limit of the lattice model, we find that the elastic scattering channel for $B=0,2$ qualitatively resembles what is observed in the U(1) case. Nonetheless, when we allow for the scattering of equal states with different momenta, we observe that the final state is highly peaked around the classical trajectory of the more energetic state, while the slower state's wavefunction becomes highly delocalized in space. For $B=1$, where the initial states are distinct, we find that the wavefunctions of the incoming bound states get entangled, forming a \textit{collective} state for the times we were able to simulate. We characterize the dynamics of these different scattering processes by computing both local observables and measures sensitive to the build of correlations in the system.

The paper is organized as follows. We first introduce the Kogut-Susskind formulation of SU(2) lattice gauge theory in (1+1)D and discuss the properties of the corresponding spin model at strong coupling in section~\ref{sec:su2_lgt}. We then describe the real-time scattering protocol employed in section~\ref{sec:scattering}, and the respective numerical results are discussed in section~\ref{sec:results}. We summarize our findings in section~\ref{sec:conclusion}.

\section{SU(2) lattice gauge theory in (1+1)D}\label{sec:su2_lgt}
We consider an SU(2) gauge theory in $(1+1)$ dimensions coupled to fundamental fermions. This model represents the simplest non-Abelian generalization of the Schwinger model~\cite{Schwinger:1951nm}, the prototypical test-bed for quantum simulations of gauge quantum field theories in two dimensions. Like the Schwinger model, and QCD in higher dimensions, this theory exhibits a confining inter-fermionic potential determined by dimensionality, 
a non-vanishing chiral condensate, and a chiral anomaly~\cite{Schwinger:1951nm,Coleman:1976uz}. 
The key difference with respect to the $\mathrm{U}(1)$ case, which we explore here, is the presence of an additional global $\mathrm{U}(1)_B$ symmetry 
associated with the conservation of baryon number $B$. This symmetry allows for gauge-invariant, color-singlet states composed of an unequal number of quarks and antiquarks. 

The baryon number counts the imbalance between matter and antimatter and is associated with the globally conserved current
\begin{align}
    J_B^\mu(x) = \frac{1}{2}\, \bar{\psi}(x)\gamma^\mu \psi(x) = \frac{1}{2}\, J_V^\mu(x) \, ,
\end{align}
where $\psi(x)$ denotes the fermion field and $J_V^\mu$ the vector current. 
The normalization is chosen such that two fundamental fermions (a color-singlet diquark) carry $B=1$. 
Although baryon number is not an exact symmetry of the full SM---being violated by weak interactions that conserve only Baryon{-}lepton number --- it remains an exactly conserved global charge in QCD. Open questions persist about which degrees of freedom \textit{effectively} carry baryon number in confining gauge theories, 
a topic closely related to the role of gluonic topology and baryon junctions~\cite{Frenklakh:2024mgu,Rossi:1977cy,Rossi:2016szw,Kharzeev:1996sq}. 
In $(1+1)$D gauge theories, by contrast, the gauge field acts as a non-dynamical potential and cannot carry or transport baryon number. 
As a result, the baryon number remains entirely associated with the matter fields, which can consistently be viewed as carrying fractional baryon number.

In the continuum, and in the temporal gauge $A_0^a=0$, the SU(2) gauge theory coupled to fundamental fermions is described by the Hamiltonian
\begin{align}
    H_{\rm cont} = \int dx \,  \bar{\psi}(x) \, \gamma^1 \big( -i\partial_x + g {A}_1^a(x){T}^a \big) {\psi}(x) \text{ } + \text{ } m{\bar{\psi}}(x){\psi}(x) \text{ } + \text{ } \frac{1}{2} {L}^a(x) {L}^a(x) \, .
\end{align}
Here $\psi$ is a two-component spinor field
\begin{align}
    {\psi}(x) = \begin{pmatrix} {\psi}_1(x) \\ {\psi}_2(x) \end{pmatrix}\,, 
\end{align}
with mass $m$. Its adjoint is defined as ${\bar{\psi}}(x) = {\psi}^\dagger(x)\gamma^0$, and $(1,2)$ correspond to the spinor components of the fields, each carrying a color charge. The fermions couple to the (reduced) electric field \(g{L}^a(x) = -\partial_t {A}_1^a(x)\) with a coupling strength $g$ and via the color matrices $T^a$. In (1+1)D these are particularly simple since they can be written directly in term of Pauli operators: \(2T^a =  \sigma^a\) with \(a = x, y, z\). Similarly, the Dirac matrices can also be written in terms of Pauli operators; we take the choice \(\gamma^0 = \sigma^z\) and \(\gamma^1 = i\sigma^y\), which respects the Dirac anti-commutation relations \(\{\gamma^\mu, \gamma^\nu\} = 2\eta^{\mu\nu}\). Finally, the gauge field can be fully integrated out by imposing Gauss's law at the level of the Hamiltonian. This constrains the set of physical states to be those satisfying $G^a |\psi \rangle = 0$, i.e. chargeless under the gauge group, where the local Gauss operator is defined as $G^a(x)= \partial_x L^a- \psi^\dagger(x) T^a \psi(x) $.

Having defined the continuum theory, we obtain a lattice formulation following the Kogut–Susskind prescription~\cite{Kogut:1974ag,Susskind:1976jm}, see also~\cite{Atas_2021}. In this representation, the continuum matter field $\psi(x)$ is mapped to discrete operators $\phi_n / \sqrt{a}$, each corresponding to a two-component spinor $\phi_n = (\phi^1_{n}, \phi^2_{n})^T$, where the superscript labels the color. Notice that the spinor components of the original field are staggered on the lattice, with the upper components seating on the even sites and the lower component occupying the odd sites. The (staggered) lattice spacing is $a$ and we consider a lattice with $N$ sites. The gauge interaction is introduced through a link operator ${U}_n \in$ SU(2) that connects sites $n$ and $n+1$. This operator can be written as ${U}_n = \exp(-i\,\Omega_n^a T^a)$ where $\Omega_n^b = -a gA_1^b$ parametrizes the gauge potential on the link . Physically, ${U}_n$ acts as a parallel transporter that mediates the interaction between the color degrees of freedom of neighboring fermions, ensuring the local gauge invariance of the theory. Under these considerations, the continuum theory can be directly mapped to the lattice theory Hamiltonian
\begin{equation}
    {H}_l = \frac{1}{2a} \sum_{n=1}^{N-1} \big( {\phi}^\dagger_n \, {U}_n \, {\phi}_{n+1}+ \text{h.c.} \big) + m \sum_{n=1}^{N} (-1)^n {\phi}^\dagger_n {\phi}_n + \frac{a g^2}{2} \sum_{n=1}^{N-1} {L}^2_n \, .
\end{equation}
Equivalently, Gauss's law can be written to enforce conservation of color charge at each site. To this end, we introduce the local operator
\begin{equation}\label{eq:Gan}
    {G}^a_n = {L}^a_n - {R}^a_{n-1} - {Q}^a_n\,,
\end{equation}
where ${Q}^a_n = {\phi}^\dagger_n T^a {\phi}_n$ is the color charge operator at site $n$. The $L_n$ and $R_n$ are the gauge link operators acting on the left and right of site $n$, respectively, and have to be distinguished due to the non-Abelian nature of the model. Equation~\eqref{eq:Gan} ensures that, when acting a state, the net flux of the electric field entering and leaving each site equals the color charge present there, thus preserving the local gauge invariance of the theory.

As mentioned above, the gauge field acts solely as a potential and can be explicitly integrated out. Firstly, we use the residual guage freedom, to remove the gauge links entering the kinetic energy term in the lattice theory. This can be achieved by color rotating each site via the operator $\hat{\Theta} = \prod_{k=1}^{N-1} \exp\left( i\, \vec{\Omega}_k \cdot \sum_{m>k} {\vec{Q}}^{\,m} \right)$, where $\vec Q^m$ is the color charge vector at site $m$. The potential term $\Omega_k$ can be taken such that $\Omega^\dagger \phi^\dagger_n U_n \phi_{n+1} \Omega_n \to  \phi^\dagger_n  \phi_{n+1}$.\footnote{The same procedure can be applied in the U(1) case~\cite{Hamer:1997dx}; note that the gauge link can be decomposed as ${U}_k = \exp\left( i\, \vec{\Omega}_k \cdot \vec{T} \right)$.}
Applying this similarity transformation to the full lattice Hamiltonian, we directly find
\begin{equation}\label{eq:H_lat_fermions}
    H  =
    \frac{1}{2 a} \sum_{n=1}^{N-1} \left(  {\phi}_n^\dagger  {\phi}_{n+1} + \text{h.c.} \right)
    + m \sum_{n=1}^N (-1)^n  {\phi}_n^\dagger  {\phi}_n
    + \frac{a g^2}{2} \sum_{n=1}^{N-1}  {L}_n^2
\end{equation}
which serves as the starting point for the qubit encoding. The fermionic field $ {\phi}_n$ is a two-component spinor defined as $ {\phi}_n = (  {\phi}_n^1 \text{ , }  {\phi}_n^2 )^T$, where each component corresponds to one of the colors of the SU(2) model. In the adopted notation, it follows that $ {\phi}_n^1 =  {\psi}_{2n-1}$ and $ {\phi}_n^2 =  {\psi}_{2n}$. The staggered fermion field can be further mapped to spin variables via a Jordan-Wigner transform 
\begin{equation}
     {\psi}_n = \prod_{k < n} (-\sigma^z_k)\,  {\sigma}_n^{-}, \hspace{0.5cm}  {\psi}_n^\dagger = \prod_{k < n} (-\sigma^z_k)\,  {\sigma}_n^{+}\, .
\end{equation}
The corresponding spin Hamiltonian takes the form
\begin{align}
    H &=  -\frac{1}{2a} \sum_{n=1}^{N-1} \left[ \sigma_{2n}^+ \sigma^z_{2n} \sigma_{2n+1}^- + \sigma_{2n+1}^+ \sigma^z_{2n+1} \sigma_{2n+2}^- + \text{h.c.} \right] \nn 
    &+ ma \sum_{n=1}^N \left( \frac{(-1)^n}{2} ( \sigma^z_{2n-1} + \sigma^z_{2n} )  \right) \nn 
    &+ \frac{a  g^2}{2}\Bigg\{ \frac{3}{8} \sum_{n=1}^{N-1} (N - n) 
        \left[ 1 - 2 \sigma^z_{2n-1} - 2 \sigma^z_{2n} \right]\nn
        &+ \frac{1}{2} \sum_{n=1}^{N-2} \sum_{m=n+1}^{N-1} (N - m)  \Big( 
\sigma_{2n-1}^- \sigma_{2n}^+ \sigma_{2m-1}^+ \sigma_{2m}^- + 
\sigma_{2n-1}^+ \sigma_{2n}^- \sigma_{2m-1}^- \sigma_{2m}^+\Big) \nn
&+  \frac{1}{16}\sum_{n=1}^{N-2} \sum_{m=n+1}^{N-1} (N - m) (\sigma^z_{2n-1}-\sigma^z_{2n})(\sigma^z_{2m-1}-\sigma^z_{2m}) \Bigg\}\, .
\end{align}

\begin{figure}[htp]
    \centering
    \includegraphics[width=0.95\linewidth]{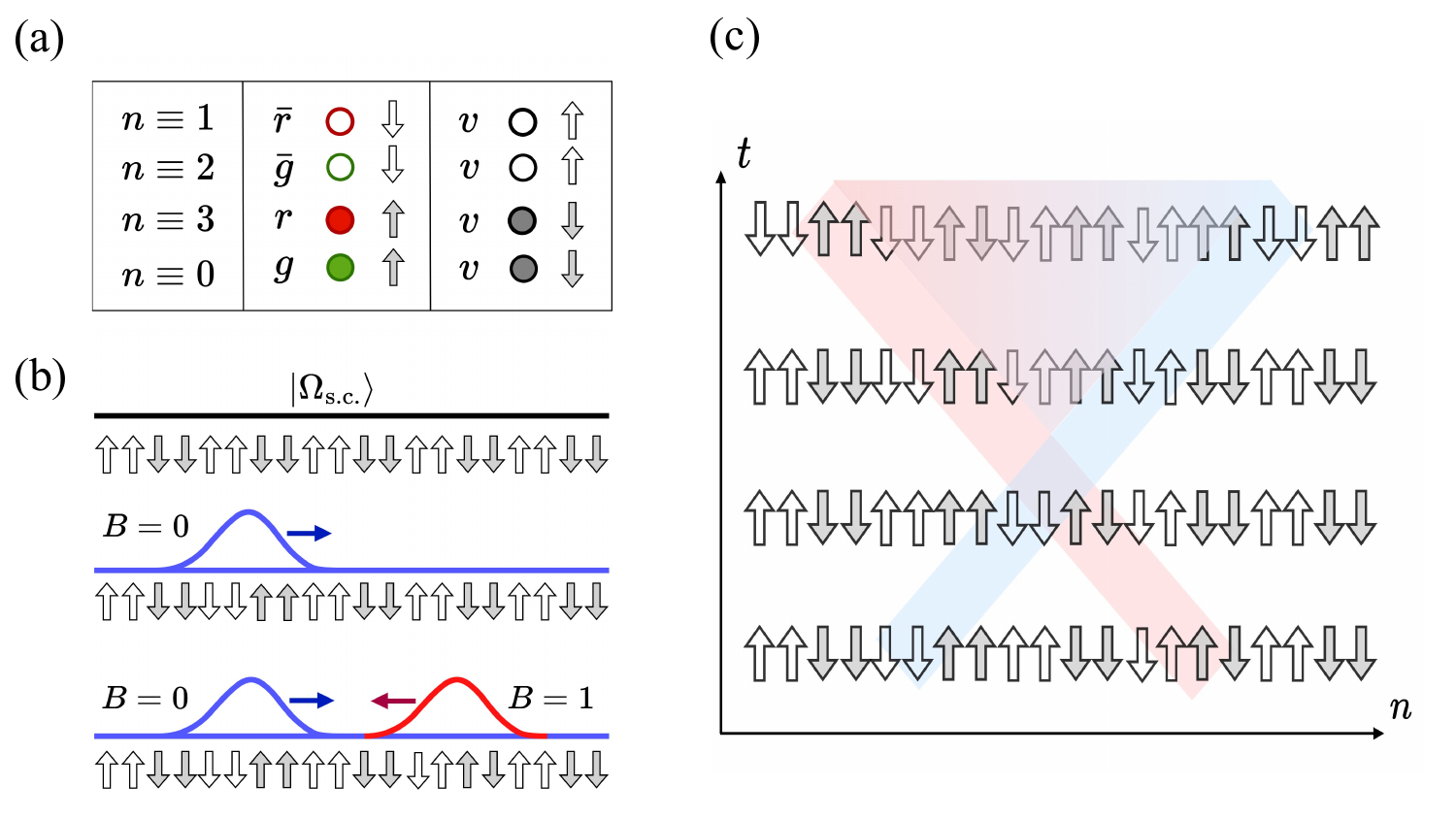}
    \caption{Schematic for the SU(2) scattering process: \textbf{(a)} qubit encoding for the SU(2) states; \textbf{(b)} state preparation for the hadron wavepackets at strong coupling; \textbf{(c)} illustration of the spacetime structure of the scattering process.
    }
    \label{fig:schematics}
\end{figure}

With this encoding, the physical content of each lattice site is determined from the residue $n \bmod 4$, considering that site numbering starts at $n = 1$, and from the local qubit state, represented by $\ket{\uparrow}$ or $\ket{\downarrow}$. According to the adopted convention, for $n \equiv 1$ the state $\ket{\uparrow}$ corresponds to the vacuum ($\mathcal{v}$) and $\ket{\downarrow}$ to a red antiparticle ($\bar{r}$); for $n \equiv 2$, $\ket{\uparrow}$ represents vacuum and $\ket{\downarrow}$ a green antiparticle ($\bar{g}$); for $n \equiv 3$, $\ket{\uparrow}$ is associated with a red particle ($r$) and $\ket{\downarrow}$ with vacuum; and for $n \equiv 0$, $\ket{\uparrow}$ corresponds to a green particle ($g$) and $\ket{\downarrow}$ to vacuum. With this assignment, any spin configuration along the chain uniquely determines the distribution of particles, antiparticles, and empty sites. The choice for the basis states is briefly summarized in Fig.~\ref{fig:schematics} (a).

As is detailed below, we shall study the theory close to its strong coupling limit, i.e. $0<m/g\ll1$ and $ga\gtrsim 1$, and it is thus convenient to briefly comment on the properties of the physical states in this regime. Since, as explained above, baryon number is a conserved global charge, it is useful to separate the particle sectors according to it. 
Considering first the $B=0$ sector, a direct analysis of Eq.~\eqref{eq:H_lat_fermions} immediately indicates that the vacuum must take the form $|\Omega_{\rm s.c.}\rangle = \ket{\uparrow\uparrow\downarrow\downarrow\cdots \uparrow\uparrow\downarrow\downarrow}$. The first excited state with only two fermionic operators corresponds to a meson bound state of a quark and an anti-quark. Its mass at leading power reads
\begin{align}
    M_m \approx 2m + \frac{3ag^2}{8} \, .
\end{align}
Indeed, this is the SU(2) analog of the Schwinger boson; higher excitations analogous to what is found in the U(1) spectrum can be similarly constructed. Interestingly, a different $B=0$ configuration can also be constructed — a gauge-invariant diquark--antidiquark ($B\bar B$) pair — which competes with the meson at strong coupling; see~\cite{Atas:2022dqm} for the analogous case in SU(3). If both clusters are local (on the same site), this state does not require gauge links, and its leading contribution to the mass is simply $M_{B\bar B}\approx4 m$, where higher-order terms are suppressed by kinetic corrections $\mathcal{O}(1/(ag)^2)$ needed to excite either the baryon or the anti-baryon. 

Moving to the baryon (and anti-baryon) sectors, one can form the lightest states by creating a single baryon, i.e. a diquark state in a color singlet $\psi^a(x) \psi^b(x) \propto \varepsilon^{ab}$. Again, its mass reads, at leading power,
\begin{align}
    M_b \approx 2m \, .
\end{align}
One can find further excitations of the baryon by delocalizing one of the quarks to a different site; however lighter states exist with an odd number of baryon/anti-baryons. 

\begin{figure}[h!]
    \centering
    \includegraphics[width=0.6\linewidth]{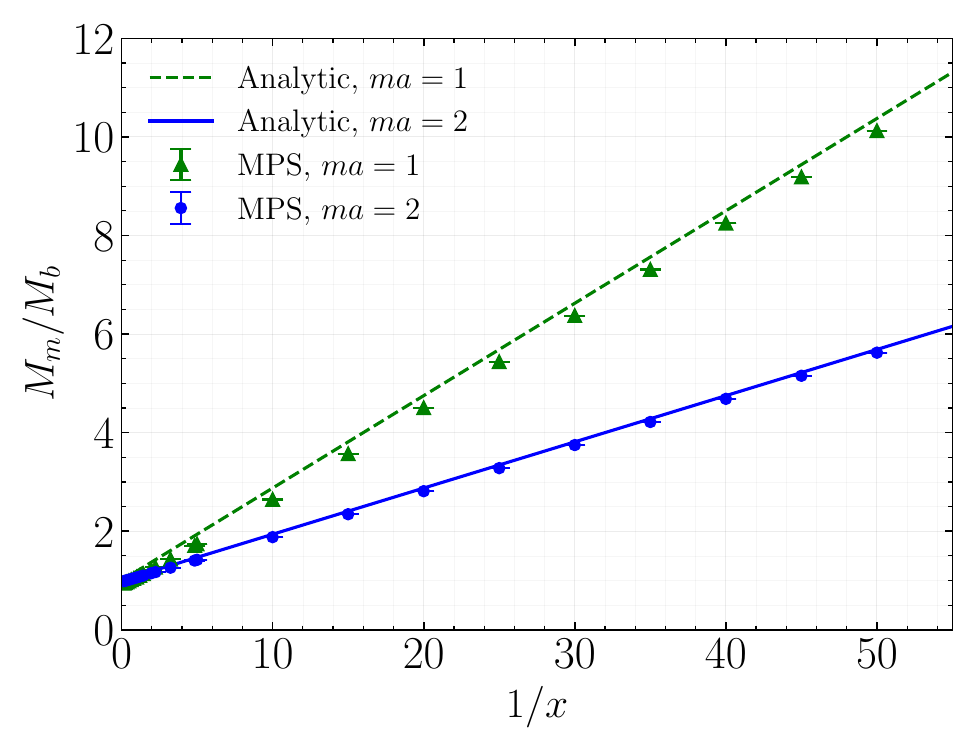}
    \caption{Meson to baryon mass ratio at evaluated with $N=60$ qubits compared with the analytical strong coupling estimation. Here $x = 1/(ga)^2$ and the strong coupling limit corresponds to $1/x\gg 1$. Uncertainty bar is computed from the absolute different between simulation results using $\chi_\mathrm{max}=80$ and $\chi_\mathrm{max}=200$.
    }
    \label{fig:Mm_o_Mb}
\end{figure}

As a result of these considerations, one has that, at strong coupling, $M_m/M_b \approx 1+ 3(ga)^2/(16 ma)$; this relation reasonably describes the tensor network result shown in Fig.~\ref{fig:Mm_o_Mb} by computing the energy gap of stationary wavepackets. Note that the mass degeneracy of these two states is well expected due to the global SU($2$) (Pauli--G\"{u}rsey) symmetry in the continuum (for a single quark flavor), which is broken on the lattice, see e.g.~\cite{Hamer:1981yq}. The same setup can also be used to study various properties of the baryon at zero temperature and finite density~\cite{Hayata:2023pkw}.

Finally, it is worth clarifying why in the numerical simulations of the $B=0$ sector the first excited state is the meson with a gauge link, rather than the $B\bar B$ configuration discussed above. From the strong-coupling expansion one indeed expects the two candidate lowest states to have schematic masses as presented above,
such that, parametrically, the diquark--anti-diquark state need not pay the electric flux cost associated with the meson string. However, this counting neglects important lattice effects. First, the $B\bar B$ state necessarily involves \emph{four} fermions, hence carries a bare mass contribution $4m$ that for realistic parameters can be larger than the mesonic contribution $2m+ \tfrac{3}{8} ag^2$. Second, while the meson is realized as a simple nearest-neighbour excitation with a single gauge link, the $B\bar B$ state requires delocalizing two diquarks across different sites in order to be gauge invariant, which introduces additional kinetic corrections $\mathcal{O}(1/(ag)^2)$. These effects push $M_{B\bar B}$ above $M_m$ at finite lattice spacing. Finally, from a variational perspective, the meson has a much larger overlap with the local Hilbert space basis, so it is naturally identified as the lowest $B=0$ eigenstate.

\section{Scattering simulation using tensor networks}\label{sec:scattering}

Having described the basic elements of the underlying theory, we now discuss the simulation of the scattering process using tensor network methods, in particular matrix product states (MPS). The numerical calculations are carried out using the \texttt{iTensor} MPS package~\cite{itensor}, making use of the native implementations of the density matrix renormalization group (DMRG) algorithm~\cite{White:1992zz,White:1993zza,Schollwock2005,schollwock2011density,frank-dmrg} and time-dependent variational principle (TDVP) algorithm~\cite{Haegeman:2011zz, haegeman2016unifying} for real-time evolution. 
The general simulation protocol implemented is as follows: we prepare first the vacuum state $\ket{\Omega}$ from the strong coupling vacuum state $|\Omega_{\rm s.c.}\rangle$ using the DMRG algorithm; we then apply the respective wavepacket operator $W_B(k, \sigma, c)$ to create the SU(2) Gaussian wavepackets $\ket{\psi_\mathrm{wvp}(k,B,c)}$ of baryon quantum number $B$ with positional spread $\sigma$ centered at lattice position $c$ travelling with momentum $k$ on the lattice (given below); after creating multiple such wavepacket states, we time evolve these hadronic states using the TDVP algorithm and measure the associated physical observables of interest. A summary of this protocol is presented in Fig.~\ref{fig:schematics} (b,c). 

In this work, we set up the SU(2) Hamiltonian using $ga=5, ma=0.2$ on $N=60$ qubits. This corresponds to the dimensionless lattice variables $\mu=2m/(ag^2)=0.016$ and $x=1/(ag)^2=0.04$ and volume $\mathcal{V} = N/\sqrt{x}=300$. Specifically, we use the DMRG algorithm with a total of $N_\mathrm{sweep}=2000$ sweeps up to a maximum bond dimension $\chi_\mathrm{max}=200$ and a numerical cutoff of $\epsilon_\mathrm{cutoff}=10^{-12}$. In fact, we found that the numerical results converge with using only $\chi_\mathrm{max}=80$, so we are stick with this lower bond dimension for all the results in this paper unless stated otherwise. In this way, the vacuum state $\ket{\Omega}$ is prepared sufficiently well and we end up with the final MPS ground state of bond dimension around $\chi_{\ket{\Omega}} = 42$ and with limited edge effects. Following the state preparation for the vacuum state, we also set the same maximal bond dimension $\chi_\mathrm{max}=80$ for the time evolution using TDVP, and use a time-step $a \delta t = 0.1$. Detailed numerical studies on different maximum bond dimensions and the time step are also provided in Sect.~\ref{sec:numerics}.

\begin{figure*}[ht!]
    \centering
    \subfigure[\;Dispersion relation for a single wavepacket.\label{fig:meons_k}]{\includegraphics[width=0.4\textwidth]{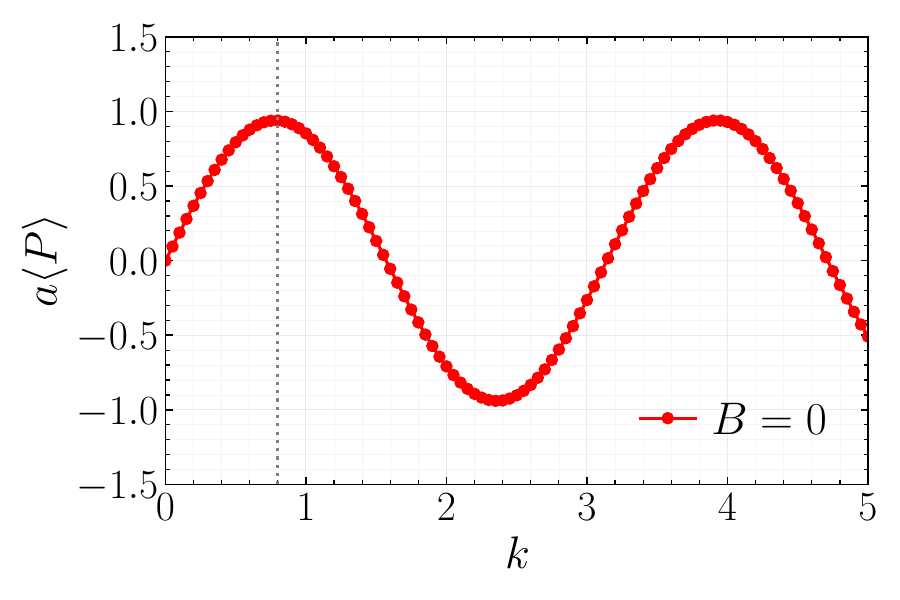}
    \includegraphics[width=0.4\textwidth]{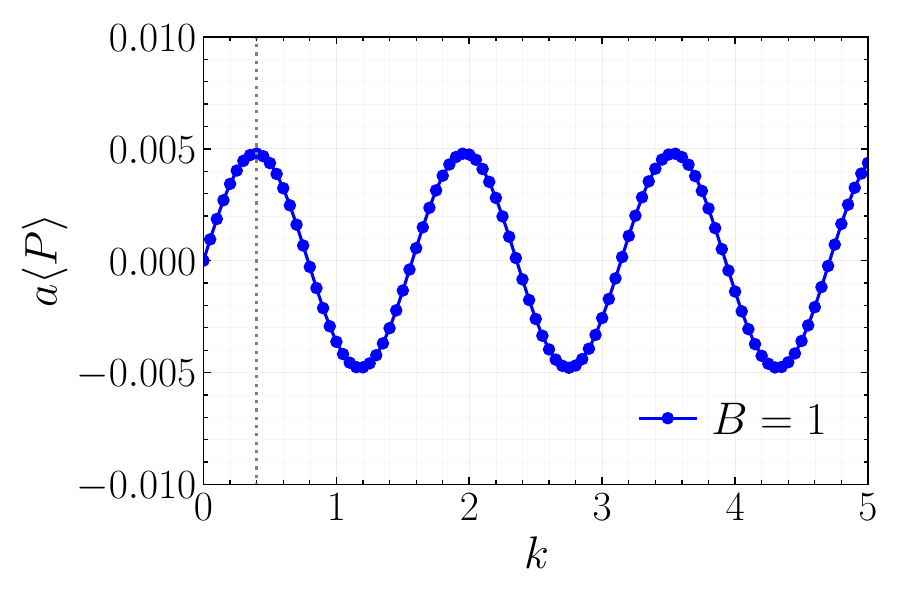}
    }
    \subfigure[\;Spatial distribution of the local baryon numbers for two initial wavepackets.\label{fig:meons_k}]{\includegraphics[width=0.4\textwidth]{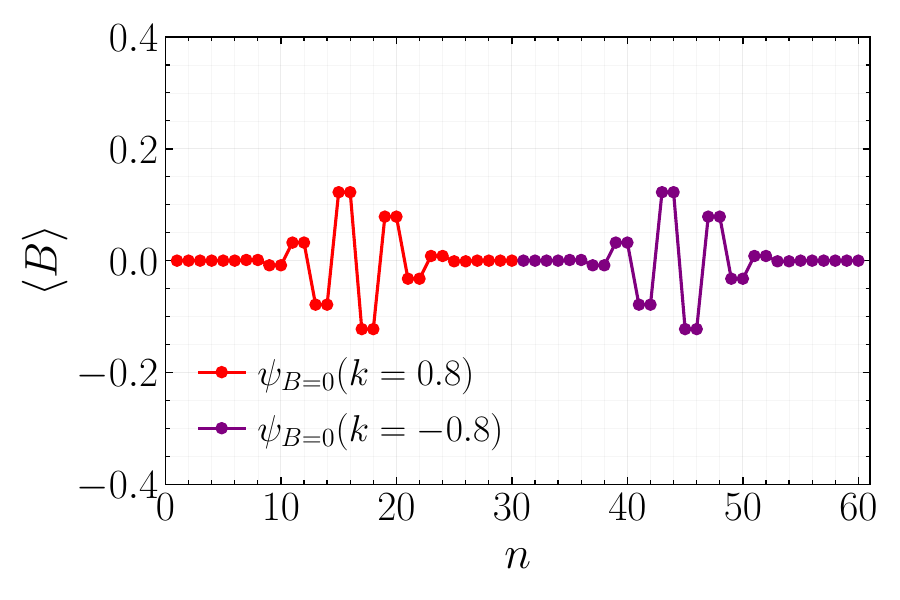}
    \includegraphics[width=0.4\textwidth]{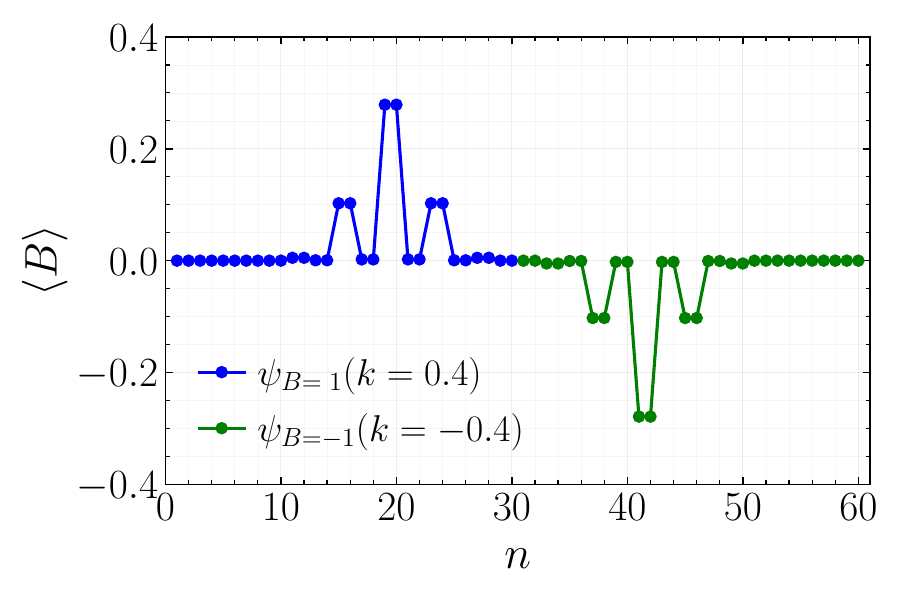}
    }
    \caption{Preparation of the wavepackets with different quantum number $B=0,-1, 1$. Momentum dependence of a single wavepacket is included. 
    }
\label{fig:wavepacket}
\end{figure*}

The wavepacket operators are explicitly defined, up to a normalizing factor, as 
\begin{align}
    \ket{\psi_B(k;\sigma,c)} \propto W_B(k; \sigma, c)\ket{\Omega} =\sum_{n=n_1}^{n_2} e^{-\frac{(n-c)^2}{2\sigma^2}} e^{ink} {\mathcal{A}}_B(n)\ket{\Omega},
\end{align}
where the local parity operators $\mathcal{A}_B(n)$ have the explicit form 
\begin{align}
    \mathcal{A}_B(n) = 
    \begin{cases}
    \sigma^+_{2n-1}\sigma^z_{2n}\sigma^-_{2n+1} - \sigma^-_{2n-1}\sigma^z_{2n}\sigma^+_{2n+1} + \sigma^+_{2n}\sigma^z_{2n+1}\sigma^-_{2n+2}  - \sigma^-_{2n}\sigma^z_{2n+1}\sigma^+_{2n+2}\;, & B = 0 \\
   \sigma^+_{2n-1}\sigma^+_{2n} \quad \text{if $n$ mod $2=0$ }\;,  & B = 1 \\
    \sigma^-_{2n-1}\sigma^-_{2n}\quad \text{if $n$ mod $2=1$ }\;,  & B = -1\, .
    \end{cases}
\end{align}
Note that this structure is inspired by the form of these states in the strong coupling limit of the lattice theory~\cite{Turco:2025jot}. They are adequate to our study as we work at small values of $x$ and $\mu$; for more discussions on the preparation of scattering wavepackets for generic model parameter choices see e.g.~\cite{Davoudi:2024wyv,Rigobello:2021fxw}. We take the wavepacket width $\sigma=4$ for all the scattering states and choose the momentum $k$ such that the expectation of the lattice conjugate momentum operator $P$ of the wavepacket is maximized. Similar to the U(1) case, the lattice pseudo-momentum operator $P$ is defined as
\begin{align}
    P = -\frac{i}{2a}\big(\sigma^+_n\sigma^z_{n+1}\sigma^z_{n+2}\sigma^z_{n+3}\sigma^-_{n+4} - \text{h.c.}\big)\,.
\end{align}
For a single wavepacket, we compute $\braket{P} = \braket{\psi_{B}(k)|P|\psi_{B}(k)}$ as a function of the momentum $k$. Figure~\ref{fig:wavepacket} shows this dependence for different $B$ values and for the initial meson and baryon preparations. As seen in the top panel, the maximal baryon momentum is much smaller than that of the meson. This arises because, in the strong-coupling regime, baryons are tightly bound diquarks whose creation operator acts locally on a single cell; their motion thus requires a correlated displacement of both quarks, suppressed by a factor $\sim (ag)^{-2}$. In contrast, mesons couple directly to the nearest-neighbor hopping term and propagate more efficiently. The bottom panel further shows that meson states, with vanishing total baryon number, display an alternating pattern of local baryon charge, while baryons (anti-baryons) carry only non-negative (non-positive) values.

To characterize the properties of the states produced after the scattering process, we consider a variety of observables capturing the formation of correlations in the system:
\begin{enumerate}
    \item \textit{Chiral condensate}: directly obtained from the latticized form of the quark bilinear
    \begin{align}
        C_n = \bar{\psi}_n \psi_n = \frac{1}{2}\big( \sigma^z_{4n-3} + \sigma^z_{4n-2} - \sigma^z_{4n-1} - \sigma^z_{4n} \big)\, .
    \end{align}
\item \textit{Baryon number}: it can be decomposed into a sum of the color content at each site, via the \textit{redness}, $\mathcal{R} = \sum_{n=1}^N \phi_n^{1\dagger}\phi_n^{1} - N/2$, and and \textit{greeness}, $\mathcal{G} = \sum_{n=1}^N \phi_n^{2\dagger}\phi_n^{2} - N/2$, operators, such that
\begin{align}
{B} = \frac{{\mathcal{R}} + {\mathcal{G}}}{2} = \frac{1}{2}\sum_{n=1}^N \phi_n^{\dagger}\phi_n- N/2 = \frac{1}{4}\sum_{n=1}^{2N} \sigma_n^z\,.
\end{align}
where the baryon number at each site is $B_n=\sigma_n^z/4$ such that $B = \sum_n B_n$.
\item \textit{Electric charge}: the non-Abelian electric charges at each lattice site $n$ have three components, which on the spin representation read
\begin{align}
    {Q}^x_n = \frac{1}{2}(\sigma^+_{2n-1}\sigma^-_{2n}+h.c.) \, ,  \, 
    {Q}^y_n = \frac{i}{2}(\sigma^-_{2n-1}\sigma^+_{2n}-h.c.)\, , \, 
    {Q}^z_n = \frac{1}{2}(\sigma^z_{2n-1}-\sigma^z_{2n})\;.
\end{align}
\item \textit{Electric energy}: the chromoelectric energy is defined from the electric term in the Hamiltonian. We compute a \textit{reduced} chromoelectric energy for each site, 
\begin{align}
    E_n \equiv \langle L^2_n \rangle = \left\langle\Big( \sum_{m\leq n} {Q}_m \Big)^2 \right\rangle
\end{align}
for each site where the color charge operator is defined as
\begin{align}
    {Q}_m^a &= \frac{1}{2} \left(
    \sigma_{2m-1}^+ \sigma_{2m}^- + \sigma_{2m-1}^- \sigma_{2m}^+, \,
    i( \sigma_{2m}^- \sigma_{2m-1}^+ - \sigma_{2m}^+ \sigma_{2m-1}^- ), 
    \,
    \frac{1}{2} ( \sigma^z_{2m-1} - \sigma^z_{2m} )
    \right)\, .
\end{align}

\item \textit{Meson and baryon projection}: to estimate the number of mesons and baryons present in the system we propose an operator that projects the meson/baryon on the lattice, acting only on subsets of sites whose parity ensures the consistency of the staggered fermion formulation,
\begin{align}
\Pi_{B} = 
    \begin{cases}
    \big\langle \Pi^{\downarrow}_n \Pi^{\downarrow}_{n+1} \Pi^{\uparrow}_{n+2} \Pi^{\uparrow}_{n+3} + \Pi^{\uparrow}_n \Pi^{\uparrow}_{n+1} \Pi^{\downarrow}_{n+2} \Pi^{\downarrow}_{n+3} \big\rangle \;\Big|_{n\,\mathrm{\,odd}}\;,  & B = 0\\
    \big\langle \Pi^{\uparrow}_n \Pi^{\uparrow}_{n+1} \big\rangle \big |_{n \; \mathrm{mod}\; 4 \;=\; 3} \;, & B = 1 \\
  \big\langle \Pi^{\uparrow}_n \Pi^{\uparrow}_{n+1} \big\rangle \big |_{n \; \mathrm{mod}\; 4 \;=\; 1} \;, & B = -1 \\
    \end{cases}
\end{align}
where $\Pi_n^{\uparrow/\downarrow}$ is the projector to the up/down spin state at the lattice site $n$. For the mesons, the internal structure of the meson composed of two quarks and two antiquarks distributed along the lattice, and therefore the operator acts on four neighbouring sites from $n$ to $n+3$ and is evaluated only for odd indices $n$. The linear combination of projectors ensures parity invariance of the observable and captures the two equivalent configurations of SU(2) along the lattice, allowing the identification of a meson symmetrically. The baryon (anti-baryon) operator is defined in an analogous way, being evaluated only on sites that satisfy $n \bmod 4 = 3 (1)$ where the operator acts on two adjacent sites, $n$ and $n+1$. This construction only gives a correct interpretation as the number of mesons and baryons in the state in the strong coupling limit of the theory; thus this is well suited for our study, see also~\cite{Ciavarella:2024lsp}. 

\item \textit{Entanglement entropy}: we consider the entropy associated to the density matrix $\rho_n$ of the subsystem formed by the lattice sites $(n,n+1)$ at time $t$, such that $S_n(t) = -\Tr \rho_n(t) \log \rho_n(t) $. This entropy can be efficiently obtained from the MPS state by computing the singular values at the external links of the subsystem.

\item \textit{Information lattice}: To characterize correlations during the scattering process, we track the flow of information across different length scales using the \emph{information lattice}~\cite{Kvorning:2021rdf,Artiaco:2024noa,Bauer:2025ljw,Artiaco:2023bot}. At each scale, we assign a \emph{local information} value  
\begin{align}
\label{eq:local_info_formula}
    i(n,\ell) &= I(\rho^\ell_n) - I(\rho^{\ell-1}_{n - 1/2}) - I(\rho^{\ell-1}_{n + 1/2}) + I(\rho^{\ell-2}_n) \, , \nn 
     I(\rho) &= \log_2[\dim(\rho)] - S(\rho) \, .
\end{align}
Here, $\rho^\ell_n$ denotes the reduced density matrix of a block of $\ell$ consecutive sites centered at position $n$ on the lattice. The indices $(n,\ell)$ thus label both the \emph{location} and the \emph{size} of the subsystem. The functional $I(\rho)$ measures the total available information in $\rho$, defined as its Hilbert space dimension minus the von Neumann entropy $S(\rho)$. By construction, $i(n,\ell)$ vanishes when $\rho^\ell_n$ contains no genuinely new correlations beyond those already present in its smaller neighboring subsystems $\rho^{\ell-1}_{n \pm 1/2}$ and $\rho^{\ell-2}_n$. Non-zero values therefore signal correlations that extend over the entire $\ell$-site block. For a recent application of the information lattice to dynamical problems in lattice gauge theories see~\cite{Artiaco:2025qqq}.

\end{enumerate}

\section{Numerical results}\label{sec:results}
In this section, we present the numerical results for the simulation of scattering in the SU(2) lattice model using MPS. Following the above discussion, we divide our results into different baryon number sectors, $B=0,1,2$. The simulation parameters for both the state preparation and real-time simulation are explicitly given in Section 3.

\subsection{$B=0$, meson-meson scattering}

\begin{figure*}[htp!]
    \centering
    \includegraphics[width=0.45\textwidth]{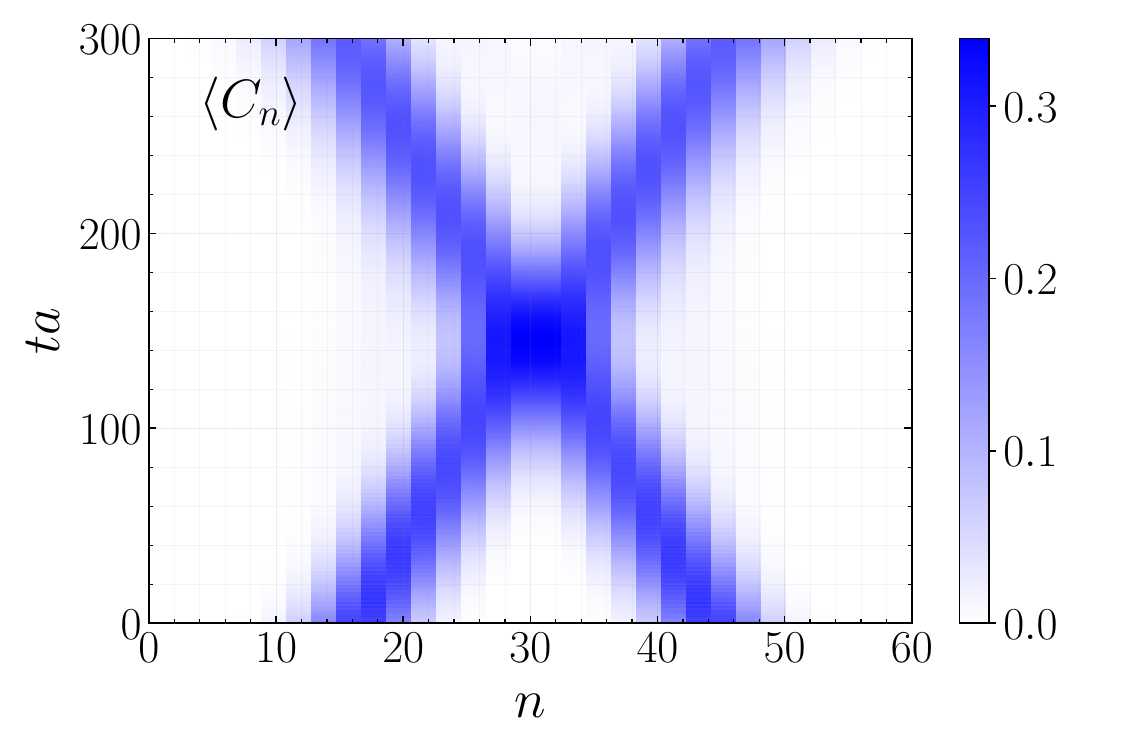}
    \includegraphics[width=0.45\textwidth]{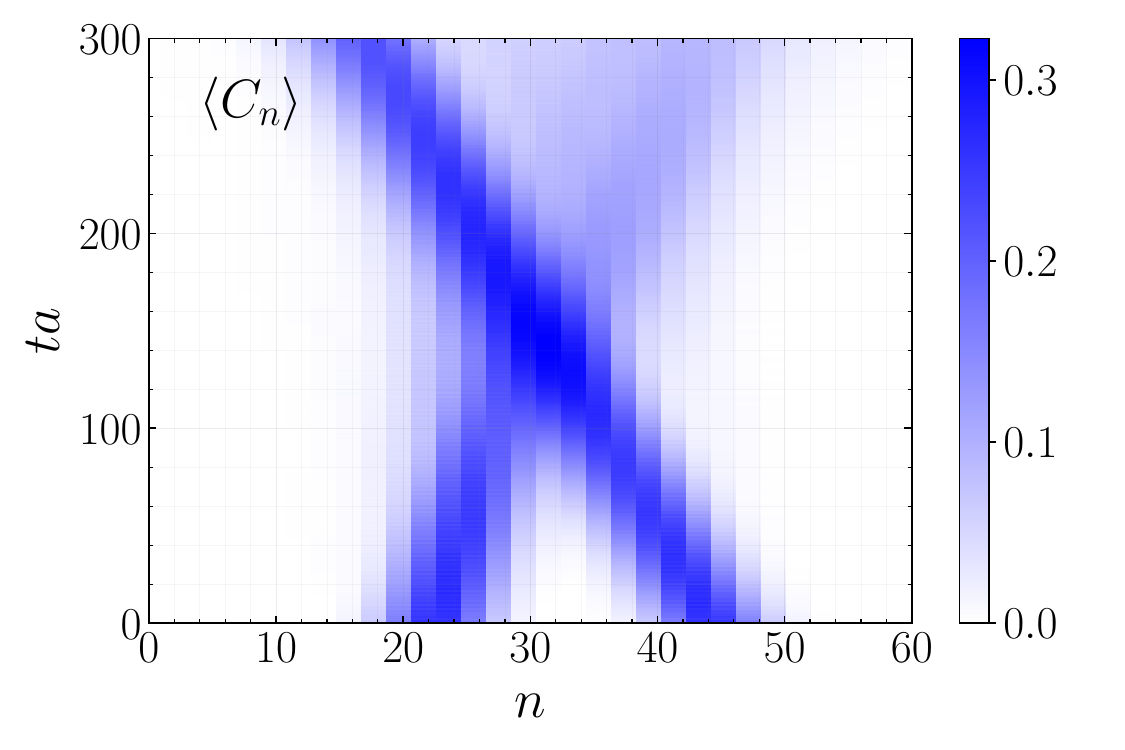}
    \includegraphics[width=0.45\textwidth]{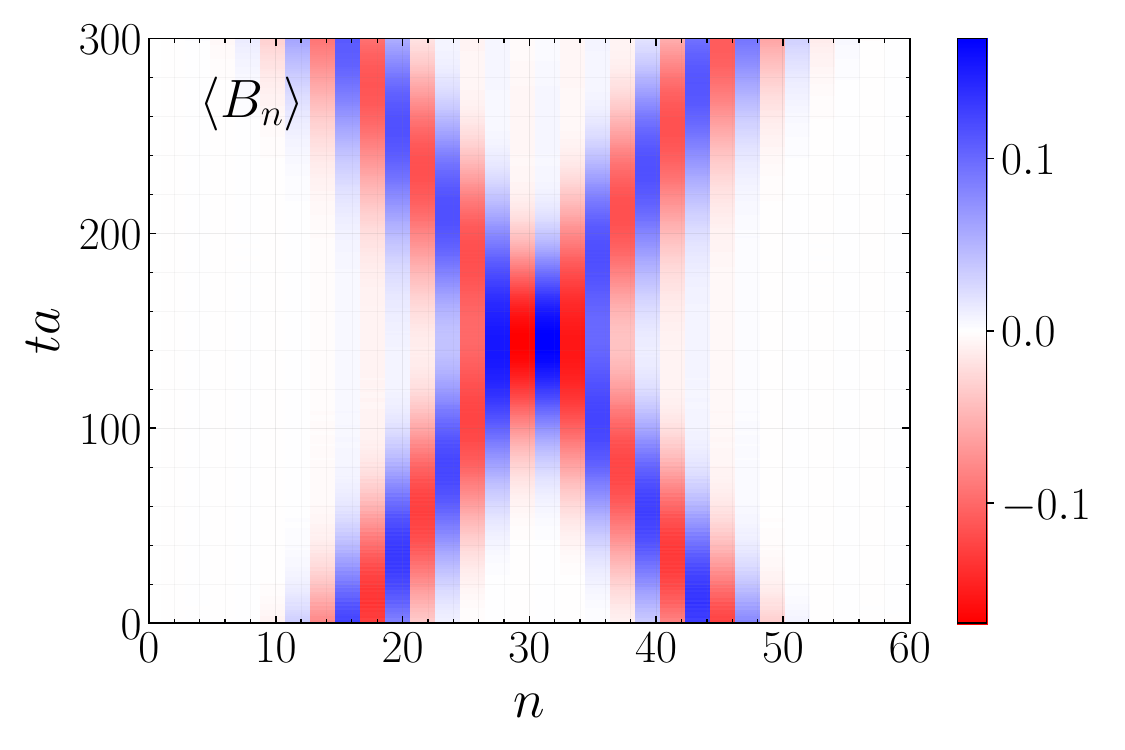}
    \includegraphics[width=0.45\textwidth]{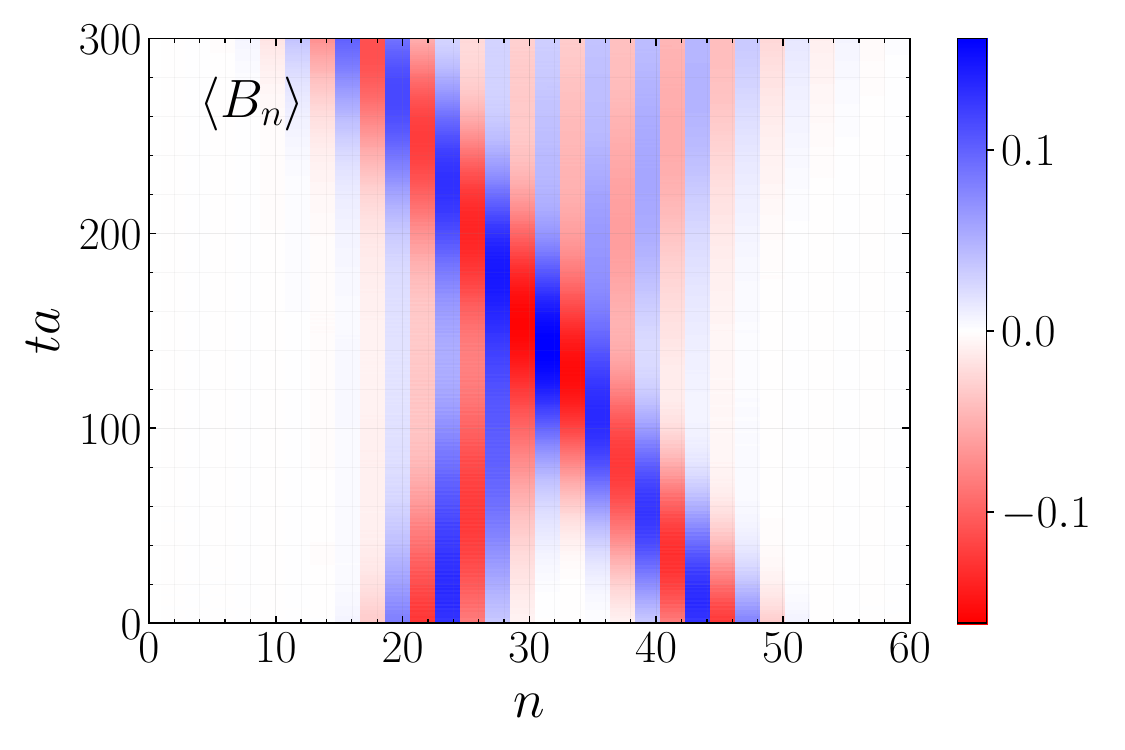}
    \includegraphics[width=0.45\textwidth]{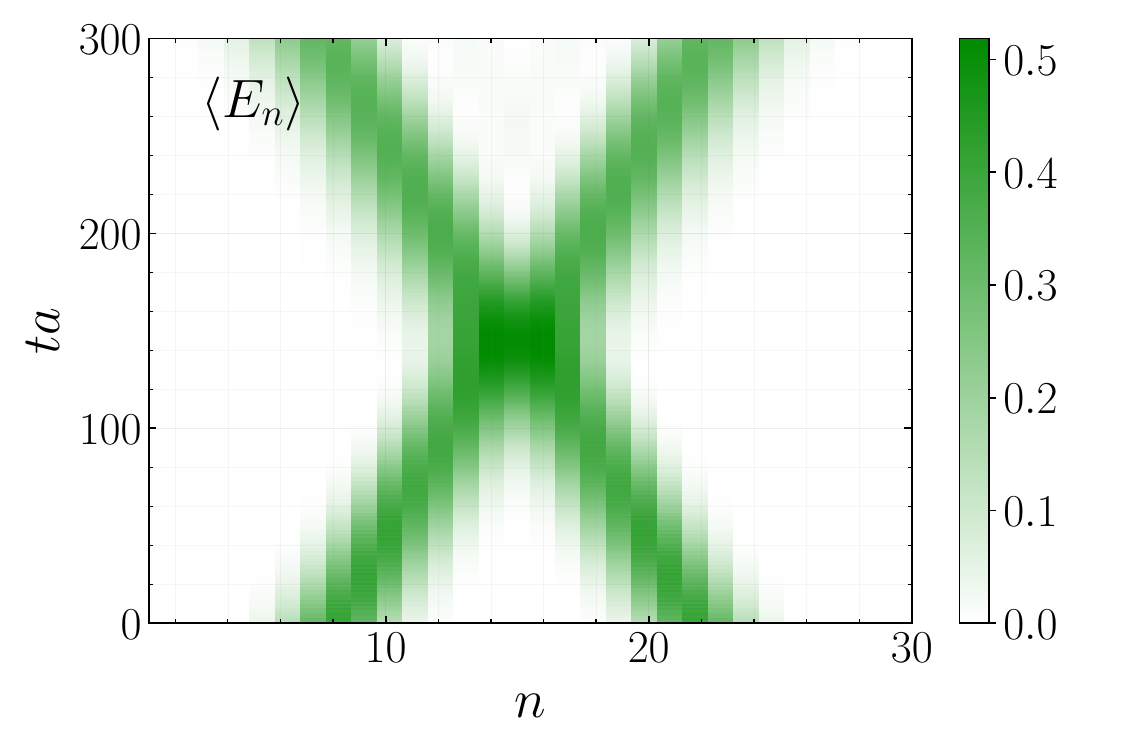}
    \includegraphics[width=0.45\textwidth]{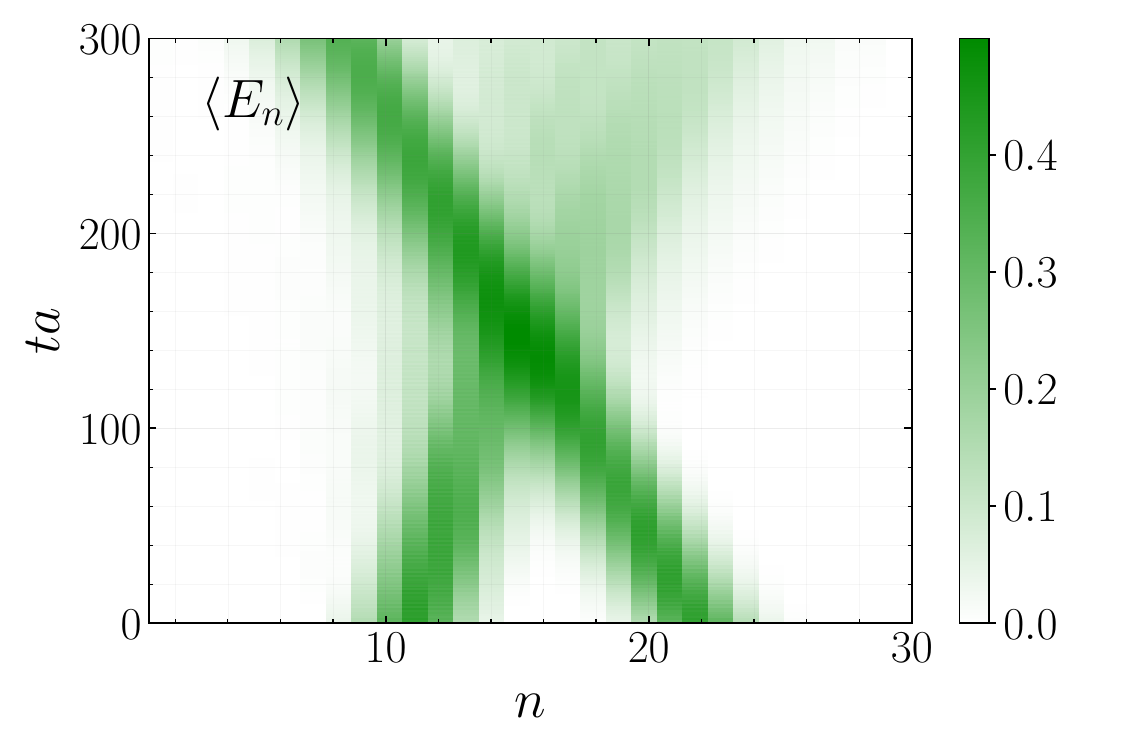}
    \caption{Condensate, baryon number and chromoelectric energy of two meson scattering using $N=60$ qubits. \textbf{Left} column shows results when both mesons are propagate at their maximal momenta; \textbf{right} column shows results for the case of initial states with different momenta. 
    }
\label{fig:mxm_results}
\end{figure*}

We first analyze scattering in the $B=0$ sector, involving two color-singlet mesons. 
The initial state is prepared by placing two meson Gaussian wavepackets with opposite momenta
centred around the middle of the chain, see Fig.~\ref{fig:wavepacket}(b). In Fig.~\ref{fig:mxm_results} we show the subsequent time evolution of the system in terms of the local condensate, baryon number and reduced electric energy. The left column corresponds to the case where the two mesons both have their maximal momenta with opposite direction, i.e, $\braket{P_l} = -\braket{P_r} = \braket{P_\mathrm{max}} =0.94a^{-1}$ with $k_l=-k_r=0.8$; the right column has the results for the case where the two mesons have different momenta, where the left meson has $a\braket{P_{l}}=0.45$ with $k_{l}=0.25$, while the right meson has $\braket{P_{r}}=-\braket{P_\mathrm{max}}$.

\begin{figure*}[htp!]
    \centering
    \includegraphics[width=0.45\textwidth]{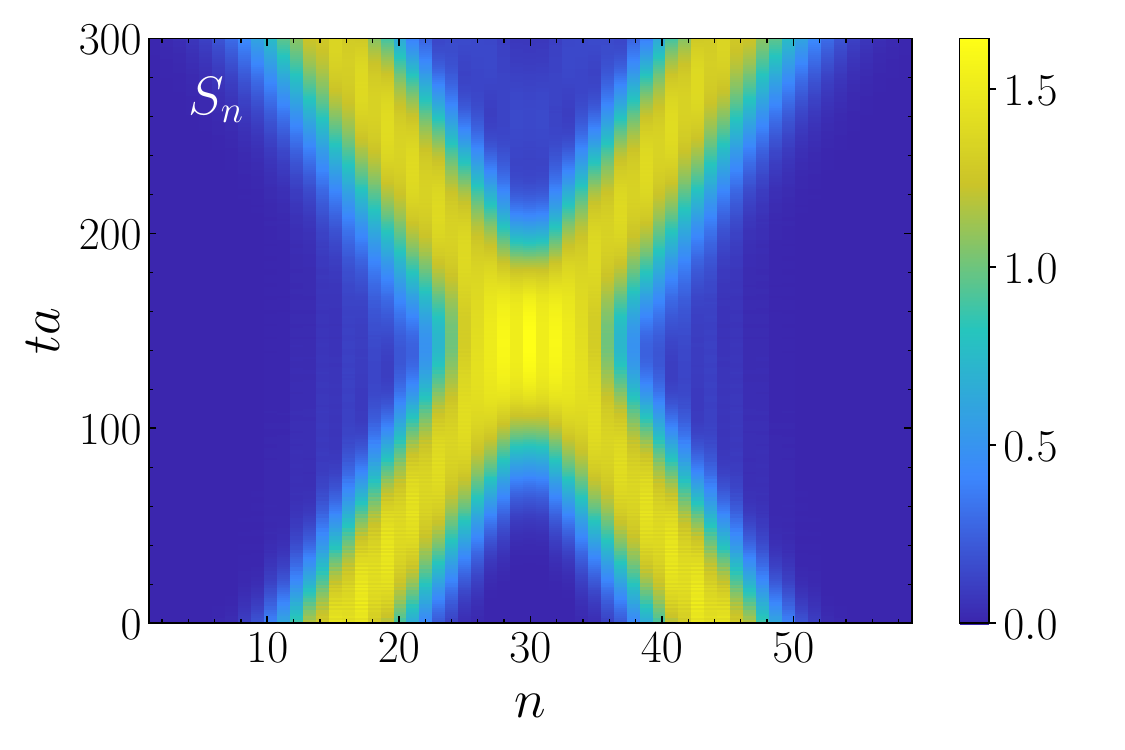}
    \includegraphics[width=0.45\textwidth]{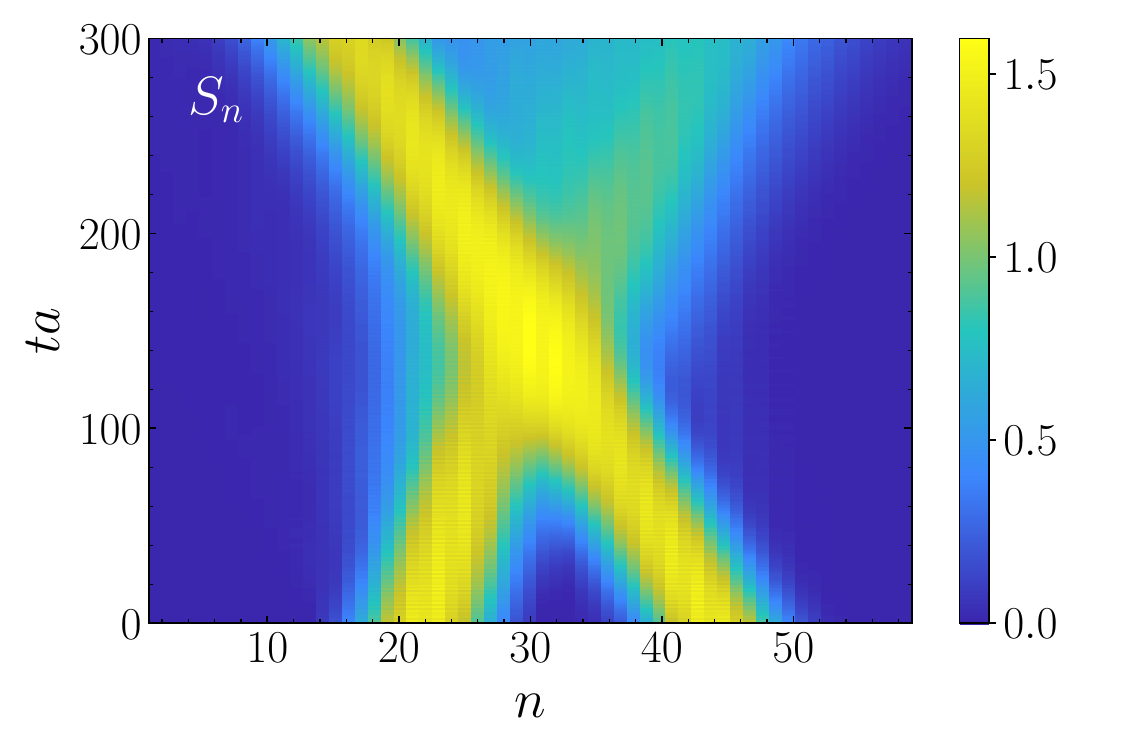}
    \makebox[\textwidth][c]{%
    \hspace{-1cm}
    \includegraphics[width=0.40\textwidth]{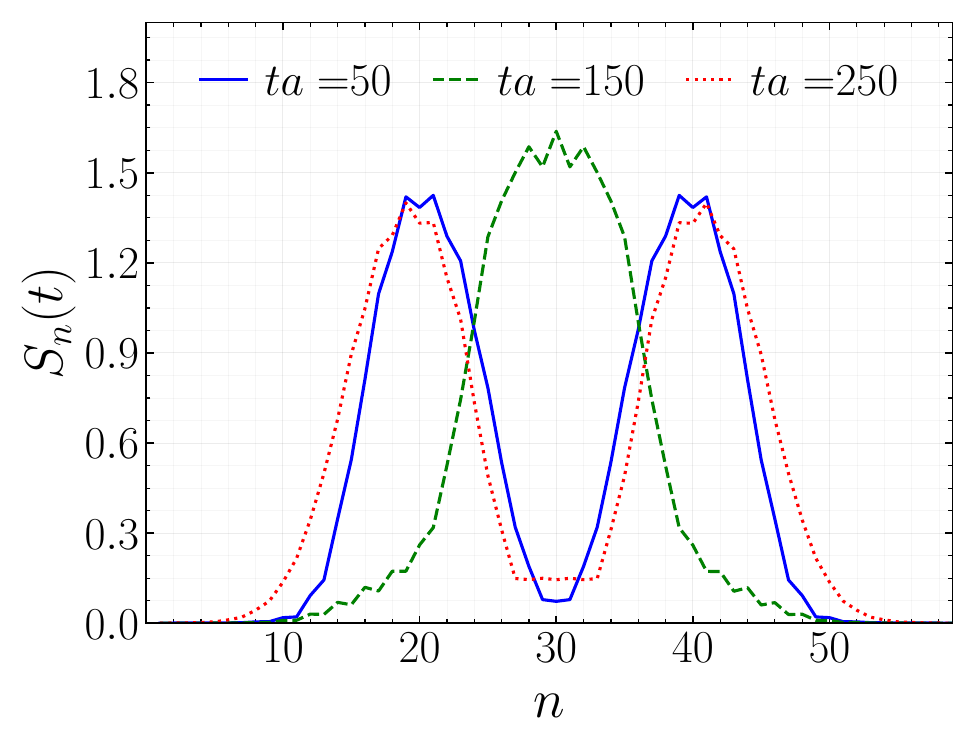}%
    \hspace{0.9cm}%
    \includegraphics[width=0.40\textwidth]{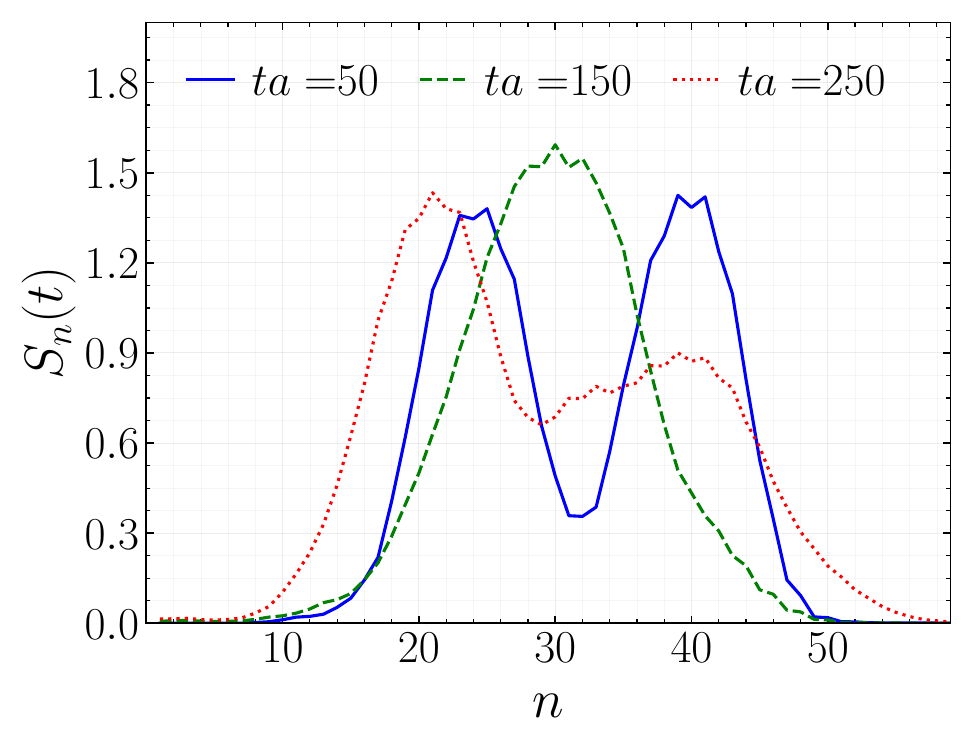}%
    }
    \includegraphics[width=0.45\textwidth]{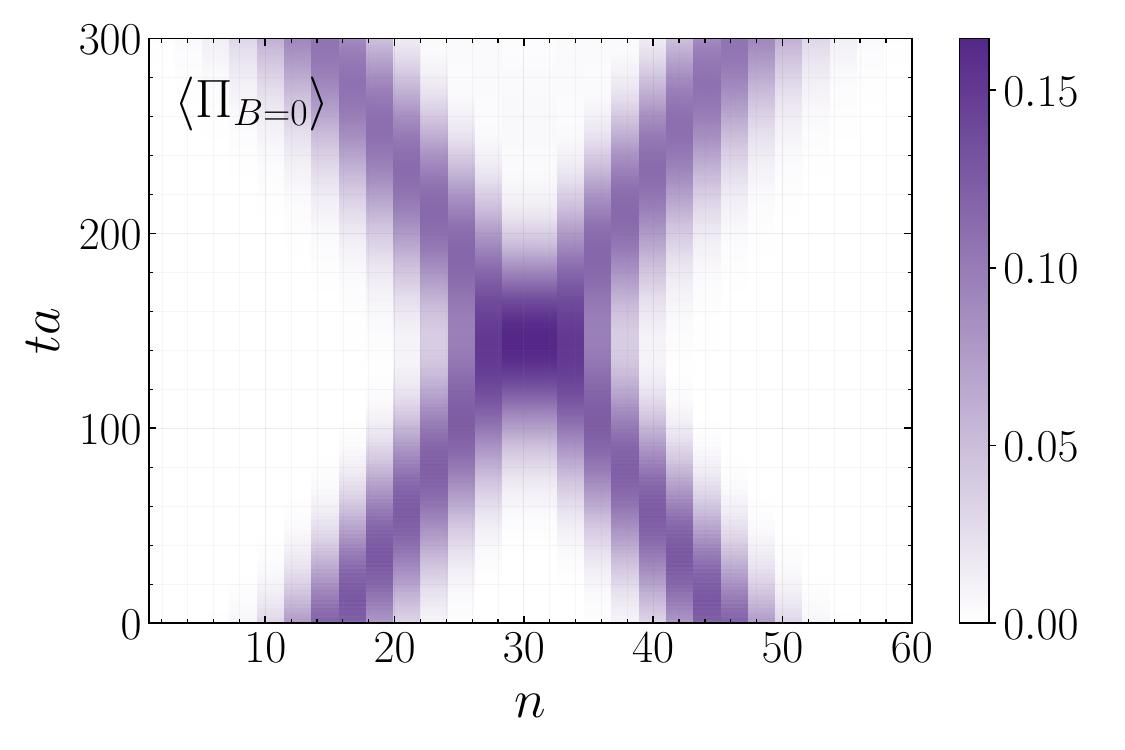}
    \includegraphics[width=0.45\textwidth]{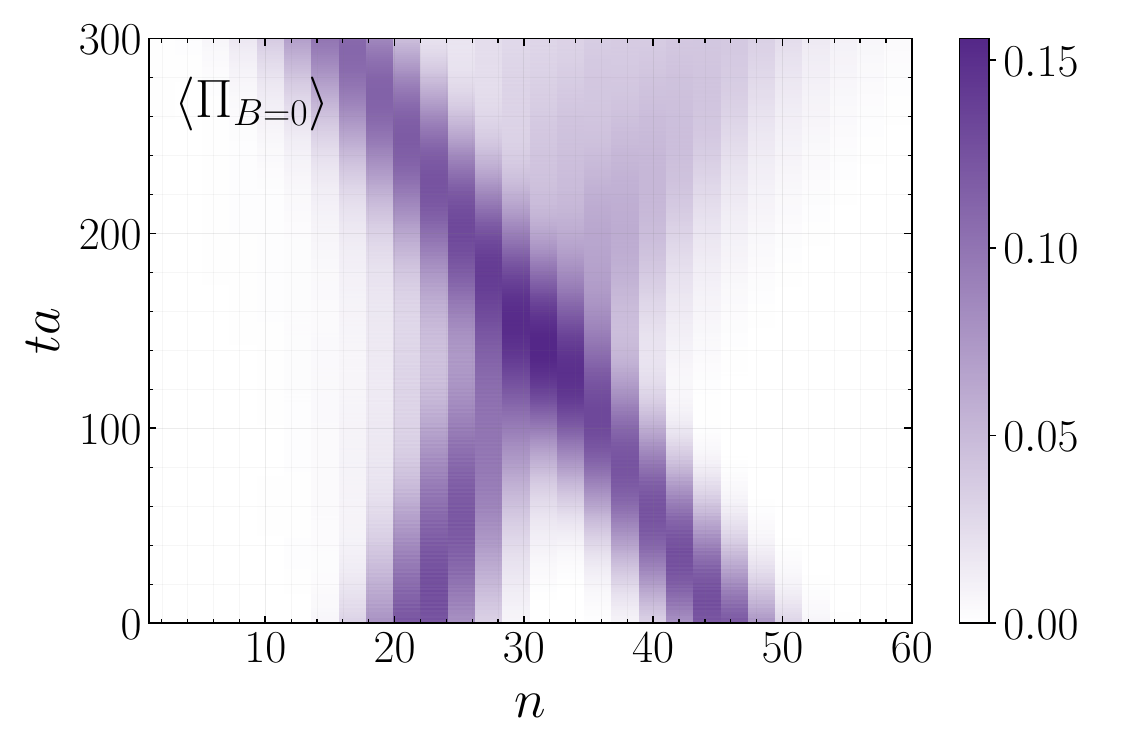}
    
    \caption{Entropy and particle projection of two mesons scattering using $N=60$ qubits. \textbf{Left} column shows results when both mesons are at maximal momenta; \textbf{right} column shows results for the case of initial states with different momenta. 
    }
\label{fig:mxm_results_ent}
\end{figure*}

\begin{figure}[h!]
    \centering
    \includegraphics[width=0.9\linewidth]{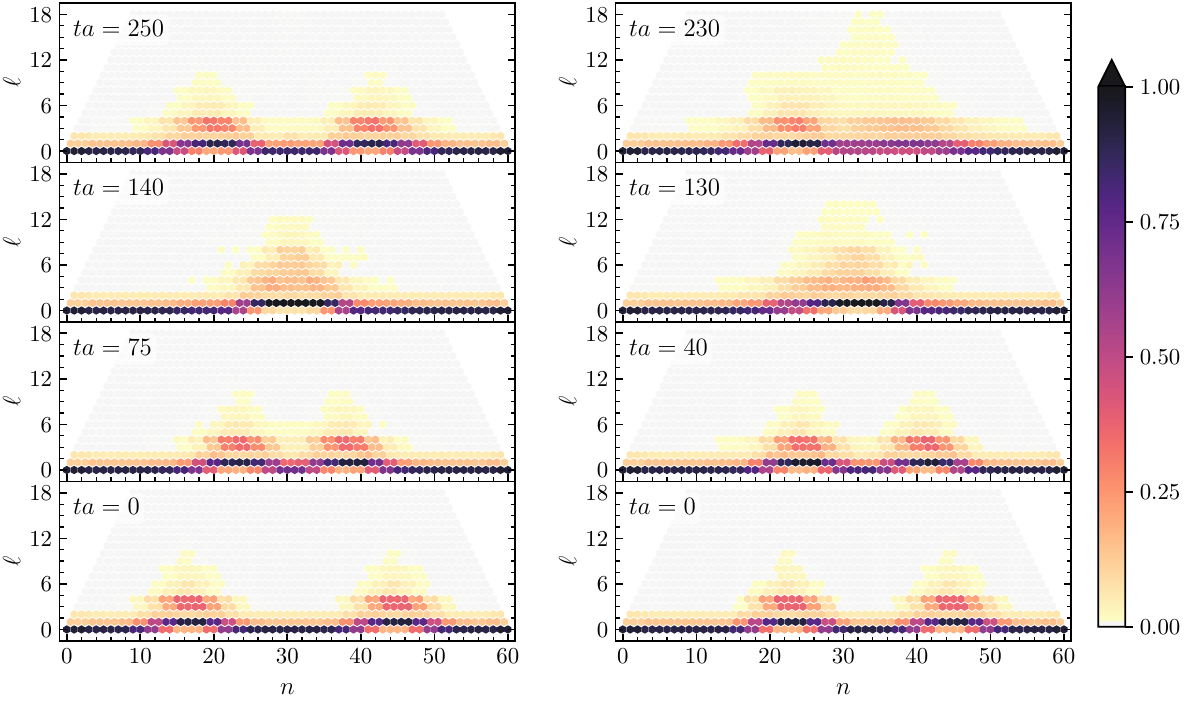}
    \caption{Local information $i(n,\ell)$ at selected times during the two meson scattering using $N=60$ qubits. \textbf{Left} column shows results when both mesons are at maximal momenta; \textbf{right} column shows results for the case of initial states with different momenta.
    }
\label{fig:mxm_results_information_lattice}
\end{figure}

Focusing on the symmetric scattering scenario, this exhibits the same behaviour as seen in the U(1) theory for processes occurring near the strong coupling limit, but below the particle production threshold~\cite{Papaefstathiou:2024zsu}. This corresponds thus to a totally elastic process where the initial meson wavepackets keep their initial structure. Indeed, in such a case one expects that the entropy of the system does not change significantly over the process, which can be seen in Fig.~\ref{fig:mxm_results_ent} (left). Here the entropy only increases when the two incoming wavepackets maximally overlap, while the final state has the same profile as the initial condition. In Fig.~\ref{fig:mxm_results_ent} (left) we show that the evolution of the entropy is directly connected to the overlap probability with the meson state, as one would expect. In Fig.~\ref{fig:mxm_results} (left) we further observe that the internal distribution of the baryon charge inside the scattering states also does not get modified during this process, again consistent with the picture of elastic scattering. Finally, in Fig.~\ref{fig:mxm_results_information_lattice} (left) we show the evolution in the information lattice: starting with two wavepackets characterized by an information peak centered around $\ell\approx 3,4$ which is mirrored at $ta=0$ and $ta=250$. However, during the interaction, the overlap of the two wavepackets leads to an increase of the information level consistent with the entropy peak.

When the mesons start with different momenta, we observe a slightly different picture in all the observables. In this case, the faster state moves ballistically as in the symmetric scattering scenario, while slower meson's wavefunction is delocalized after the scattering. Note though that the process is still elastic. This can be cleanly observed in the entropy distribution in Fig.~\ref{fig:mxm_results_ent} (right), where the entropy density is the largest along the classical direction of the faster state. In contrast, the entropy associated to the final state of the right moving state decreases significantly after the scattering, as can be seen in the second row of Fig.~\ref{fig:mxm_results_ent}.
This observation is supported in more detail by the results for the information lattice in Fig.~\ref{fig:mxm_results_information_lattice} (right), where one cleanly observes that the final state has the same information per scale characterizing the initial condition, albeit with a very broad profile for the right-moving state.

\subsection{$B=1$, meson-baryon scattering}
\begin{figure*}[htp!]
    \centering
    \includegraphics[width=0.45\textwidth]{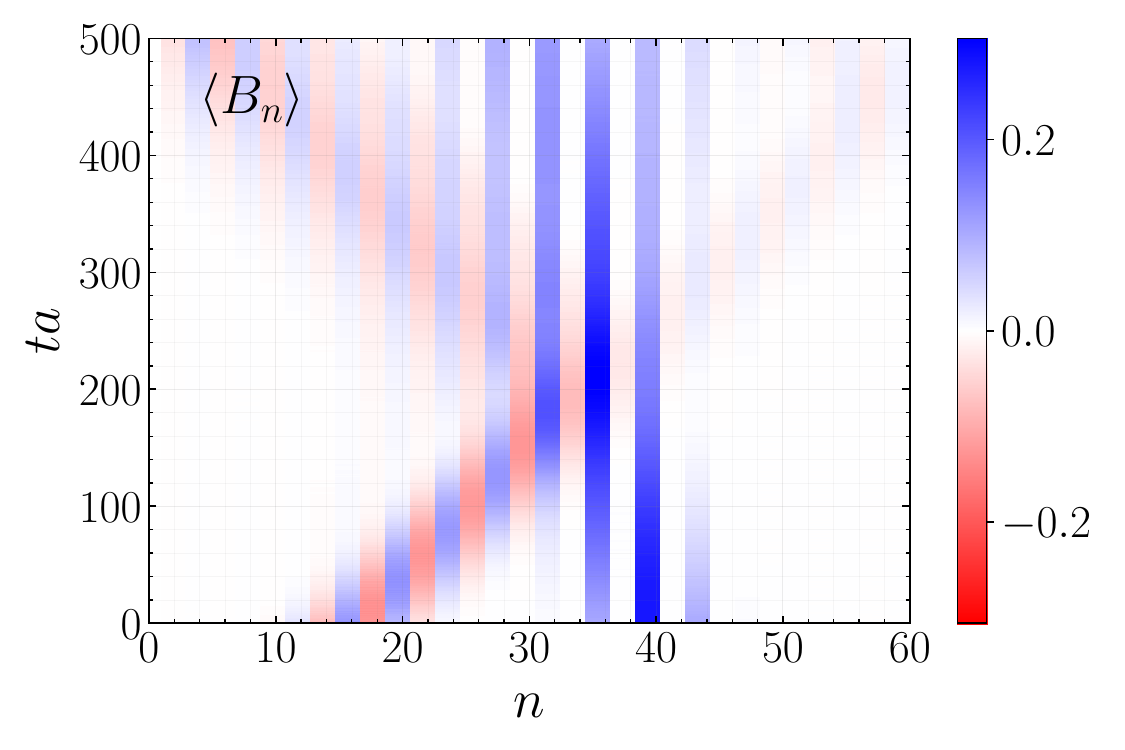}
    \includegraphics[width=0.45\textwidth]{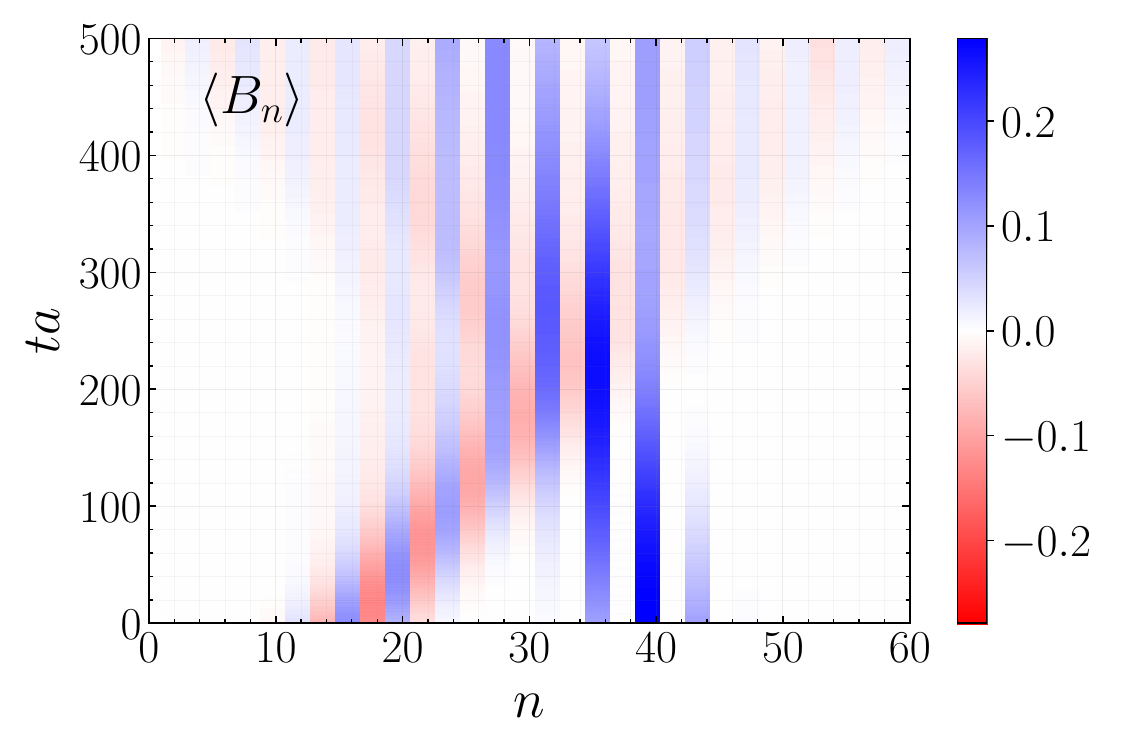}
    \includegraphics[width=0.45\textwidth]{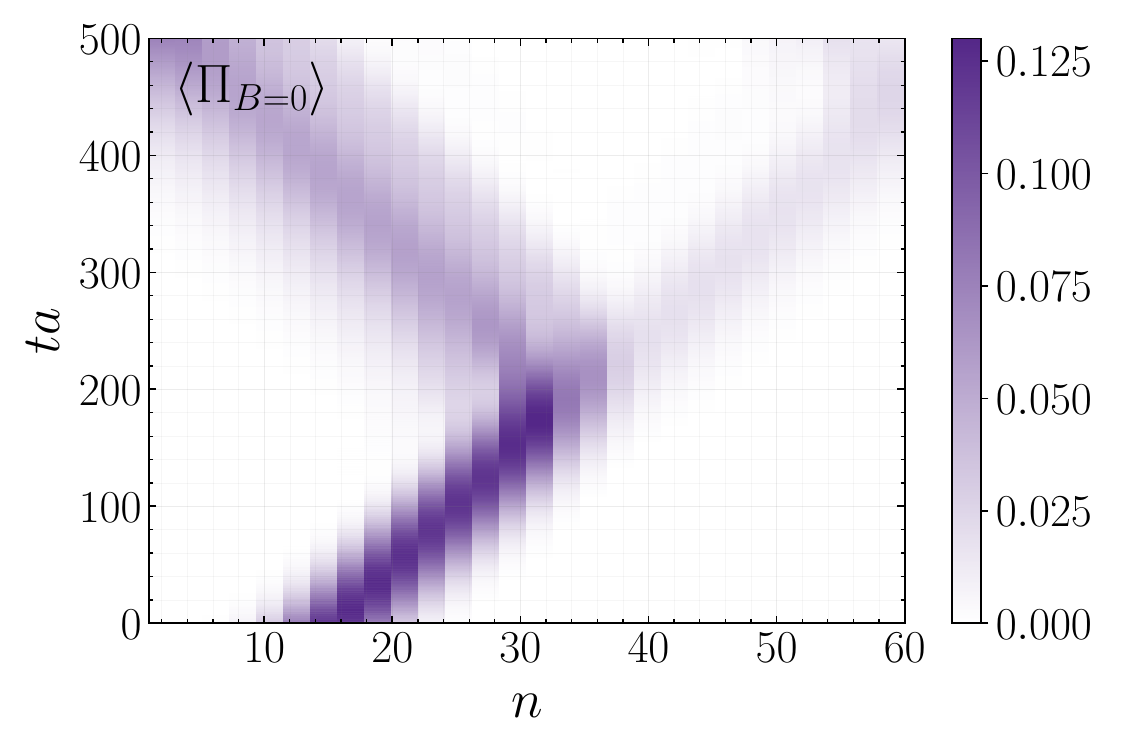}
    \includegraphics[width=0.45\textwidth]{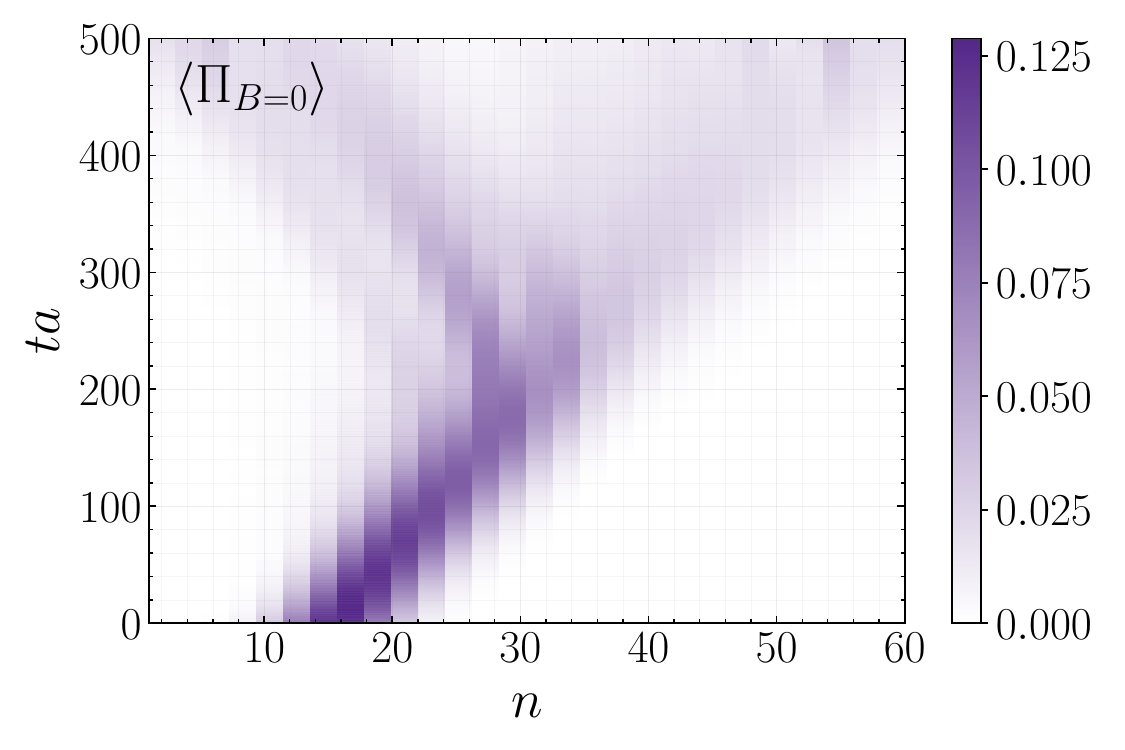}
    \includegraphics[width=0.45\textwidth]{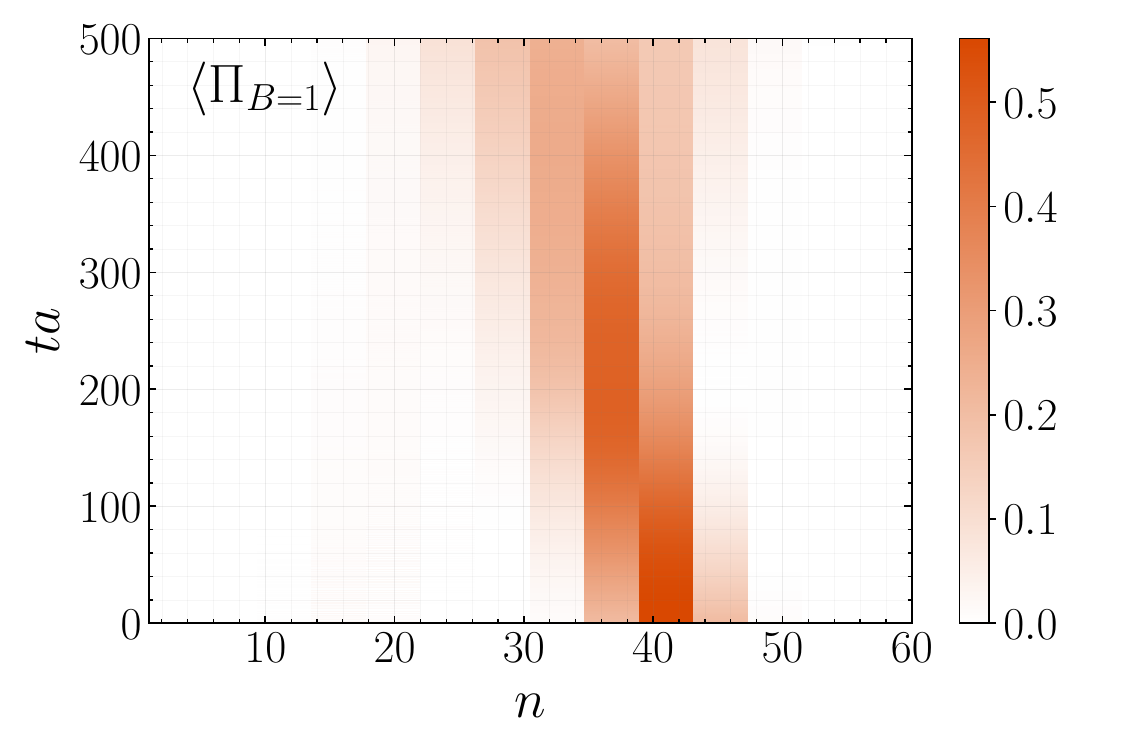}
    \includegraphics[width=0.45\textwidth]{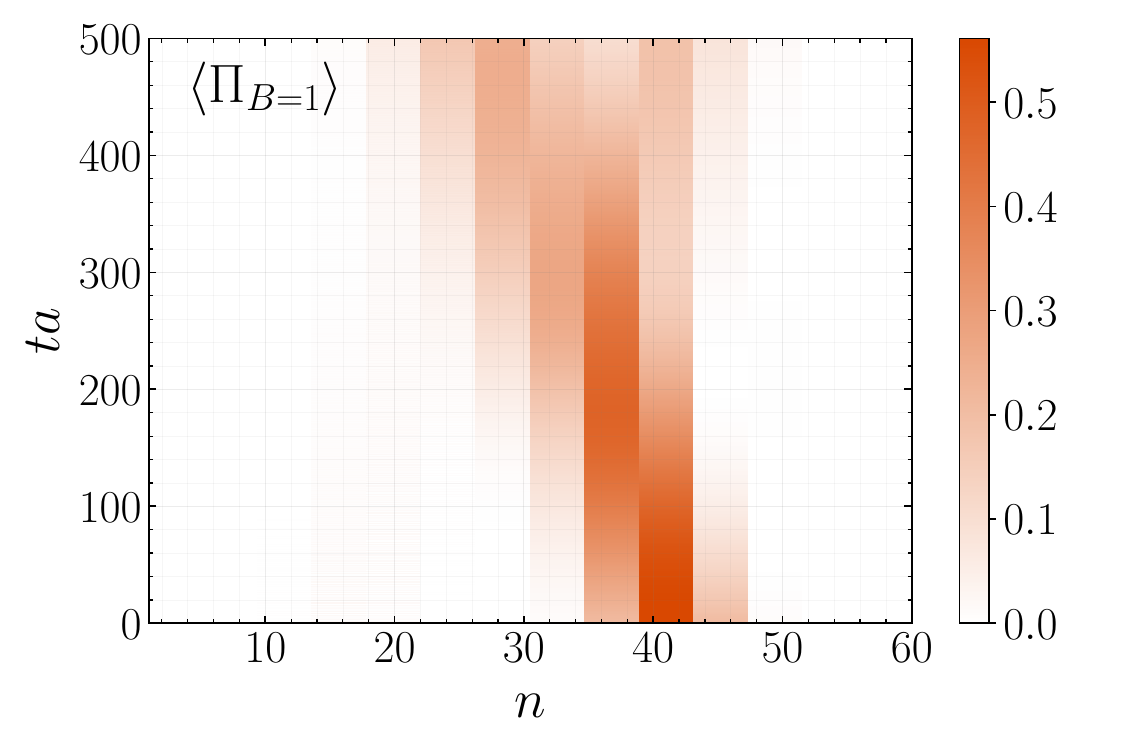}
    \includegraphics[width=0.45\textwidth]{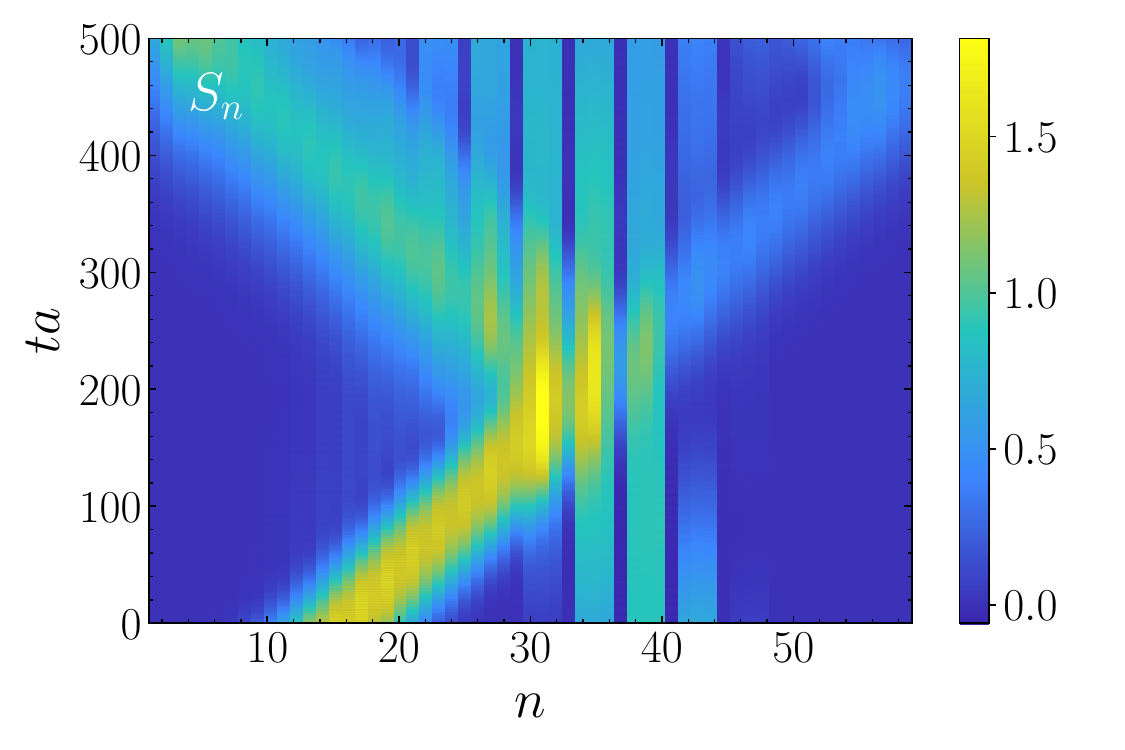}
    \includegraphics[width=0.45\textwidth]{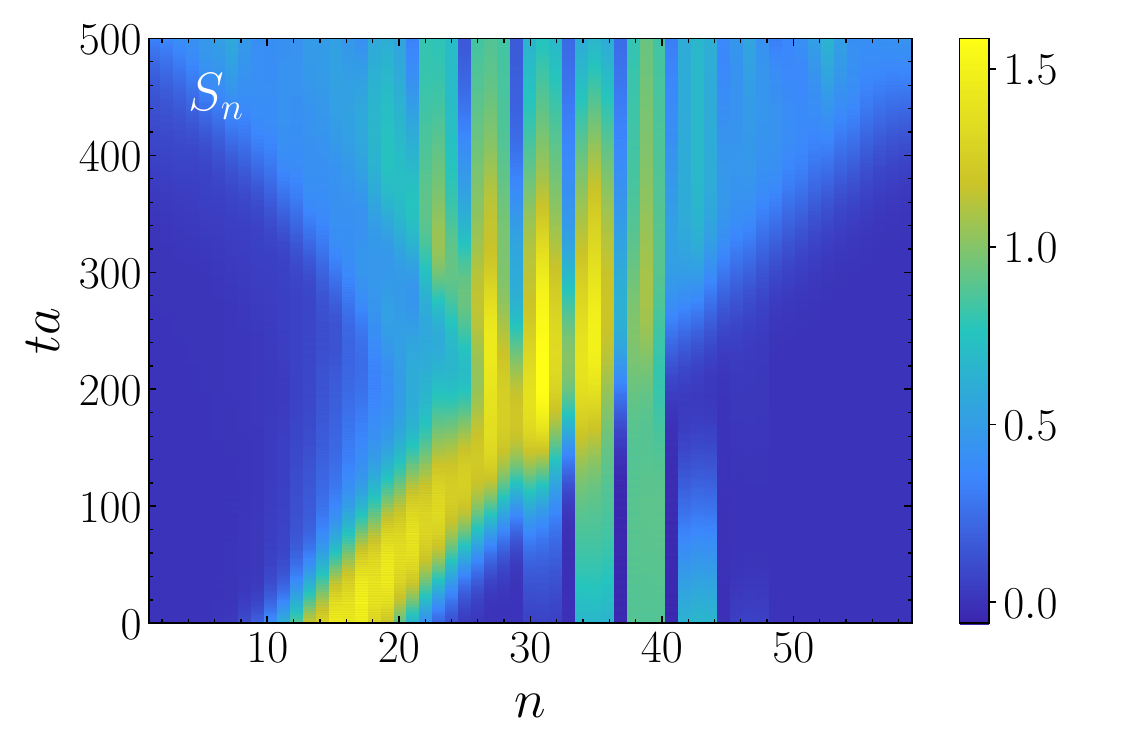}
    \caption{Baryon number, particle projection, and entropy of $B=1$ meson-baryon scattering using $N=60$ qubits. \textbf{Left} column shows results when both meson/baryon are at maximal momenta; \textbf{right} column shows results when mesons/baryon have different initial momenta.
    }
\label{fig:mxb_results}
\end{figure*}

 The $B=1$ sector reveals qualitatively new features due to the coexistence and mixing of mesonic and baryonic excitations. Here, a meson wavepacket is initialized on the left side of the lattice and propagates toward a baryon wavepacket on the right as in Fig.~\ref{fig:mxb_results}. Two representative momentum configurations are studied: (i) a fast meson on the left with $k_l=0.8$, corresponding to maximal lattice momentum $a\braket{P_l} = 0.94$, and (ii) a slower meson with $k_l=0.4$, yielding $a\braket{P_l} = 0.68$. The baryon wavepacket on the right is kept the same in both scenarios, with a maximal momentum of $a\braket{P_r}=-0.0048$ at $k_r=-0.4$.

\begin{figure}[h!]
    \centering \includegraphics[width=0.925\linewidth]{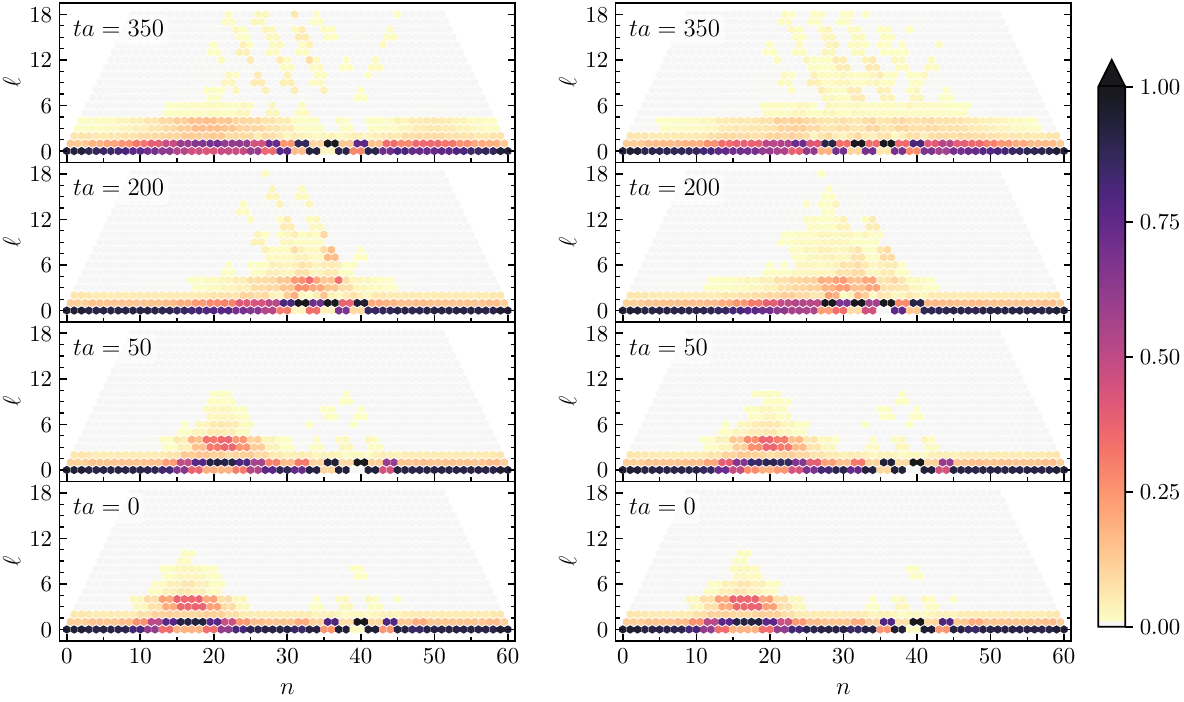}
    \caption{Local information $i(n,\ell)$ at selected times during meson-baryon scattering using $N=60$ qubits. \textbf{Left} column shows results when the right-moving meson is at its maximal momentum; \textbf{right} column shows results when the meson is moving at a slower momentum. The baryon momentum is fixed in both scenarios.
    }
\label{fig:mxb_results_information_lattice}
\end{figure}

From Fig.~\ref{fig:mxb_results} one can observe that in both scenarios the meson state is delocalized, with part of it being reflected and the remaining tunnelling over the baryon state. This is clearly seen when computing the meson and baryon projections, which show the baryon to remain mainly unchanged, while the meson probability is spread in space. More, considering the internal distribution of the baryon number inside these states, we observe that the characteristic pattern of the two sates interfere constructively. Nonetheless, the entropy evolution is rather distinct from the $B=0$ sector, where in the unequal momentum scenario, after the scattering, the entropy peak survives around the scattering point. This indicates that, for the simulated times, the two wavefunctions overlap significantly and form a single state. Note that the difference between $B=0$ and $B=1$ case is mostly connected to the fact that the scattering states are not symmetric rather than the non-Abelian nature of the of the setup.
Finally, in Fig.~\ref{fig:mxb_results_information_lattice} we show the corresponding results using the information lattice. This detailed mapping of the correlations in the system supports the previous statements: the baryon state remains localized along its classical trajectory, while the meson completely overlaps with the latter. Notice that now new information levels $\ell$ are populated (at least not significantly), indicating that there is no formation of a new state with a different correlation profile, see discussion in~\cite{Artiaco:2025qqq}.

\subsection{$B=2$, baryon-baryon scattering}
\begin{figure*}[htp!]
\centering
    \includegraphics[width=0.32\textwidth]{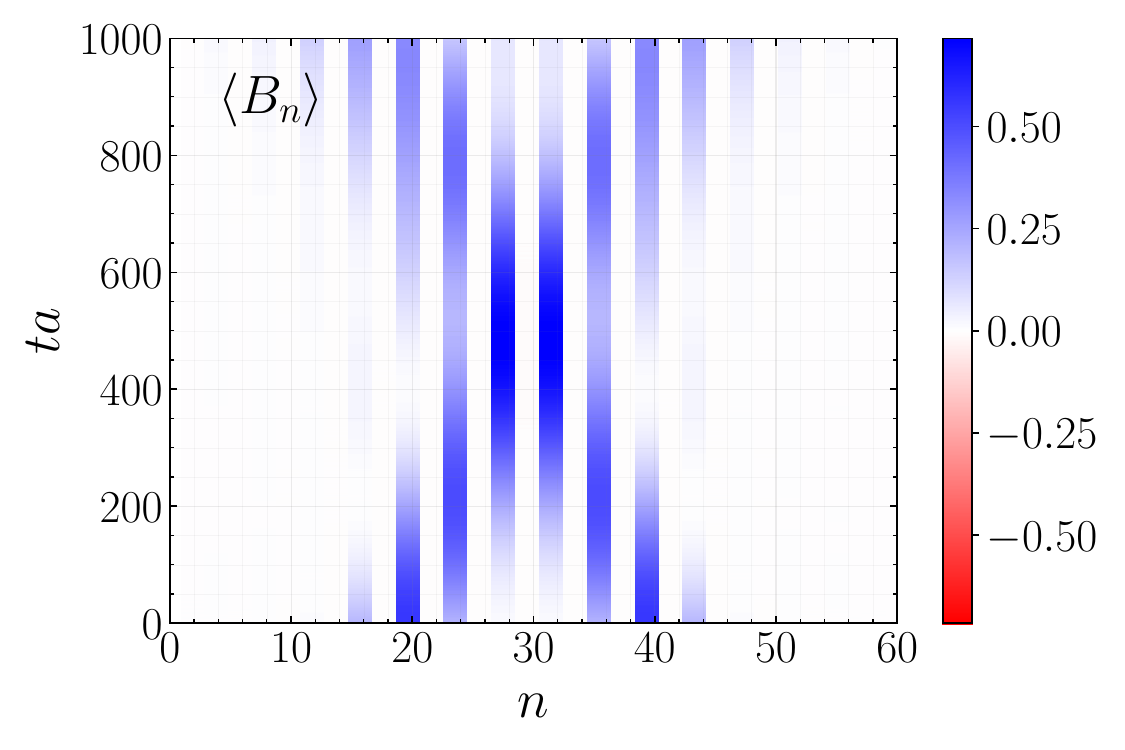}
    \includegraphics[width=0.32\textwidth]{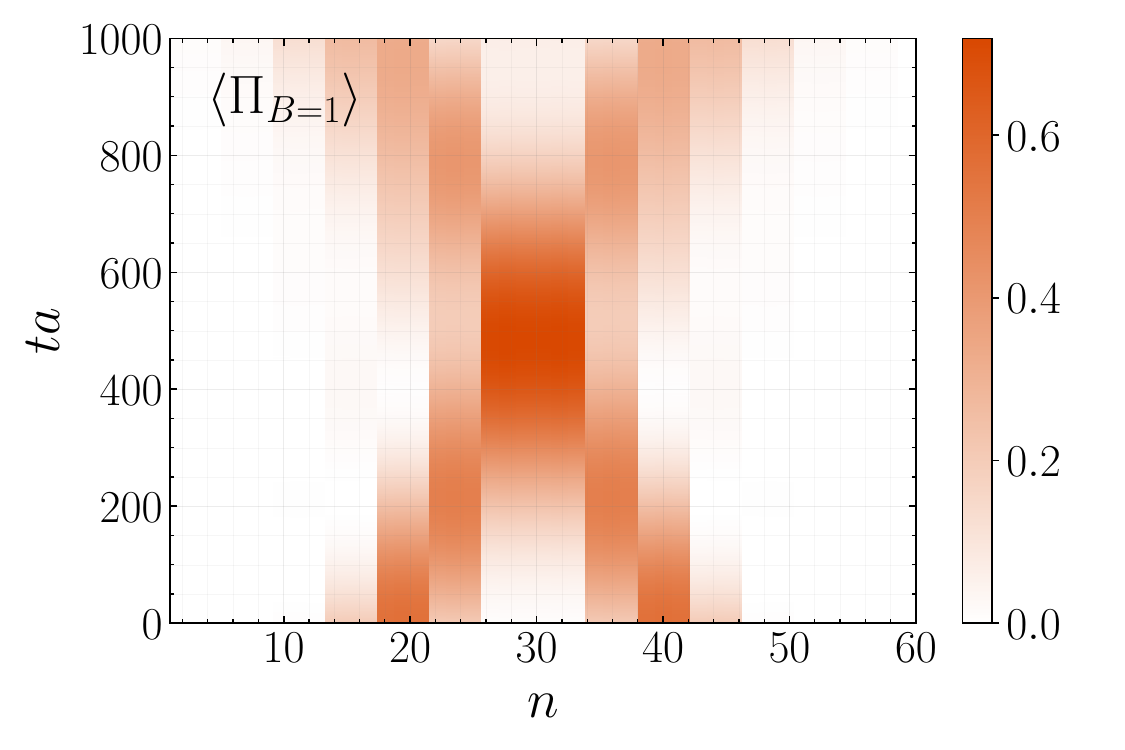}
    \includegraphics[width=0.32\textwidth]{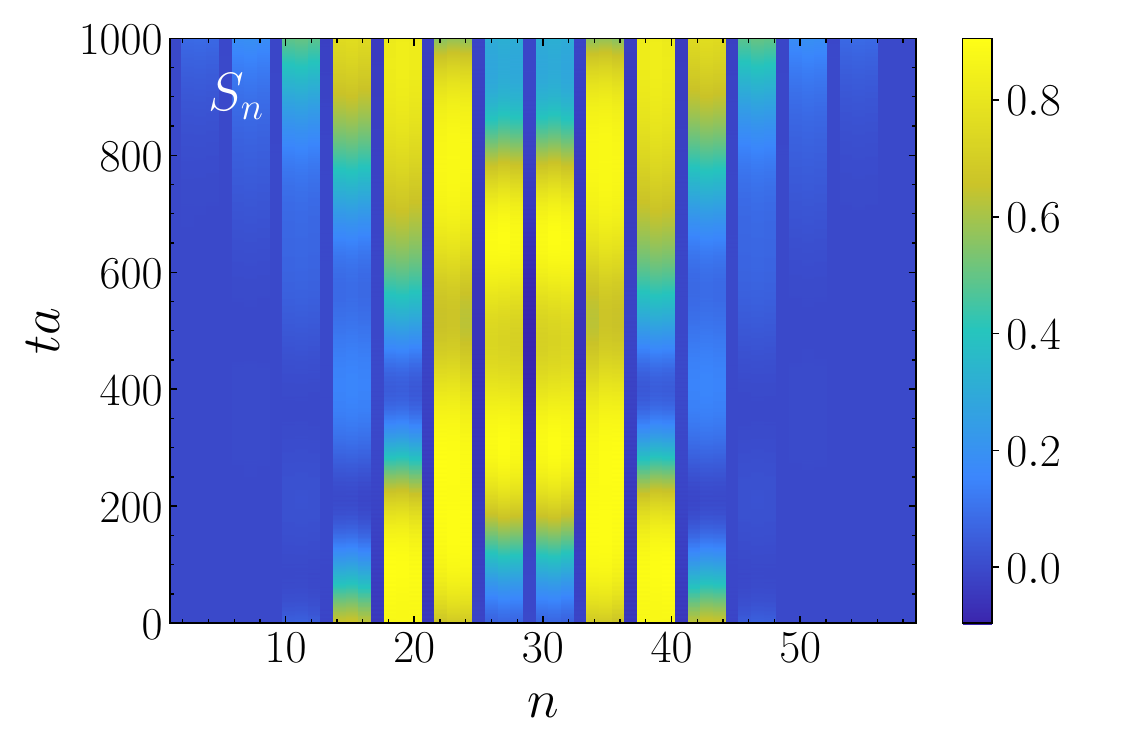}
    \caption{Baryon number, particle projection, and entropy for $B=2$ baryon-baryon scattering using $N=60$ qubits. 
    }
\label{fig:bxb_results}
\end{figure*}

Finally, we explore the $B=2$ sector, corresponding to the collision of two baryons. The initial configuration consists of two baryon wavepackets moving toward each other with equal and opposite momenta $a\braket{P}=\pm 0.0048$ at $|k|=0.4$. The process in this sector is again reminiscent of what occurs in the U(1) theory, with a perfectly elastic scattering. This is reflected in all the observables we consider; we show a selected combination of results in Fig.~\ref{fig:bxb_results}. Since in the current numerical implementation we were not able to achieve higher momentum modes for the baryons, the inelastic channels is kinematically out of reach.

\subsection{Numerical convergence}\label{sec:numerics}
In this subsection, we provide a detailed demonstration for the numerical convergence in our real-time scattering simulation result using tensor network. As we already demonstrated convergence in the meson and hadron masses (see Fig.~\ref{fig:Mm_o_Mb}), we now focus on the numerical benchmark in the real-time simulation using TDVP algorithm. In particular, we use the two-site entanglement entropy $S_n(t)$ as the main target observable to quantify the error and convergence associated with the simulation.

\begin{figure}[h!]
    \centering
    \includegraphics[width=0.45\textwidth]{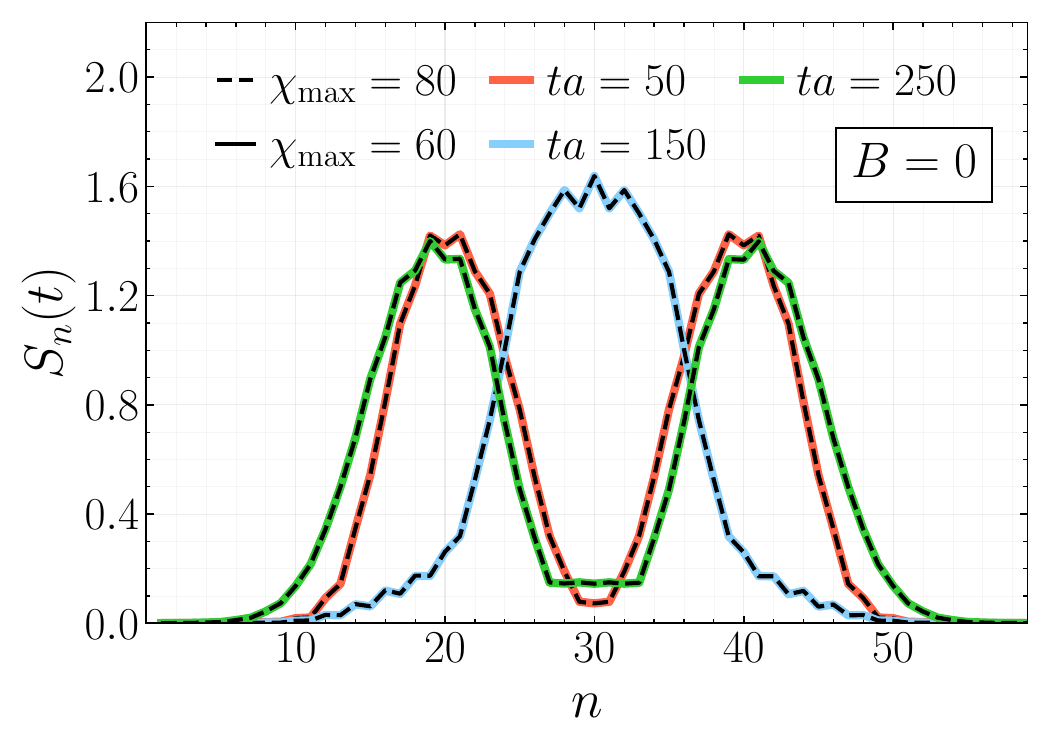}
    \includegraphics[width=0.45\textwidth]{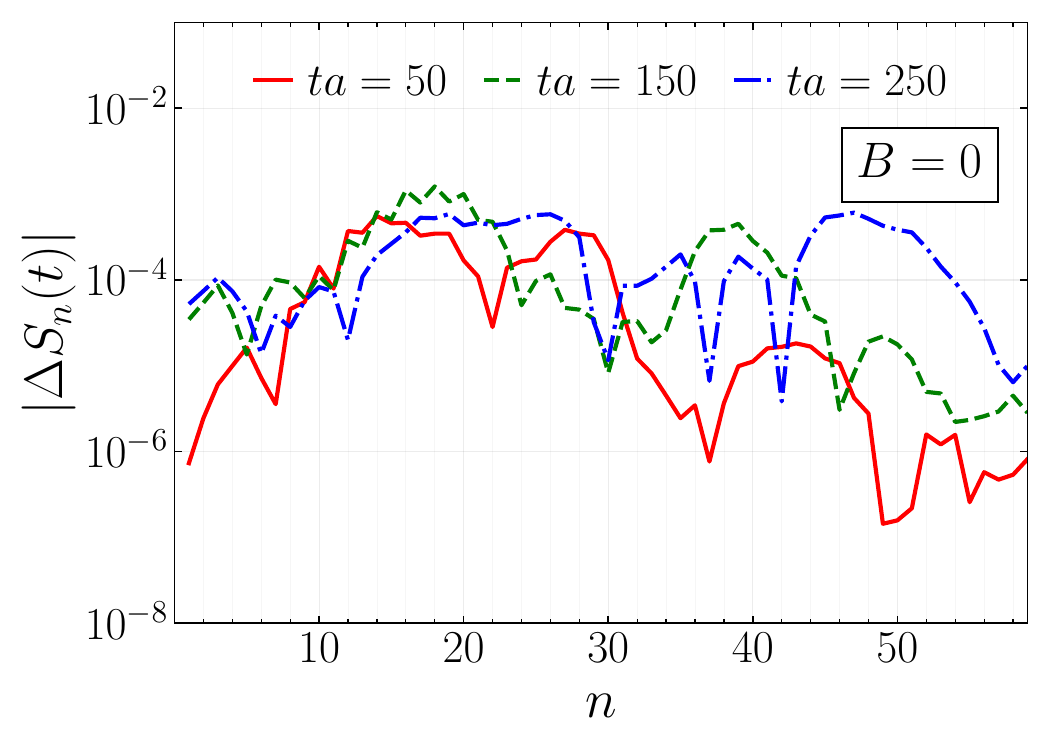}
    \includegraphics[width=0.45\textwidth]{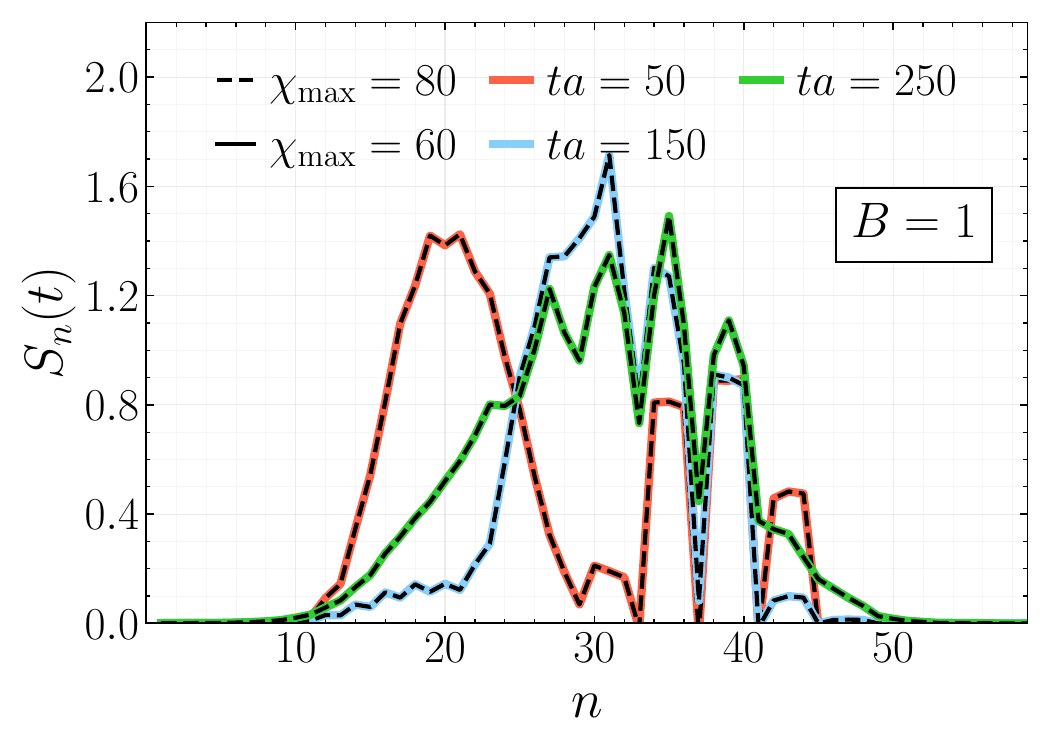}
    \includegraphics[width=0.45\textwidth]{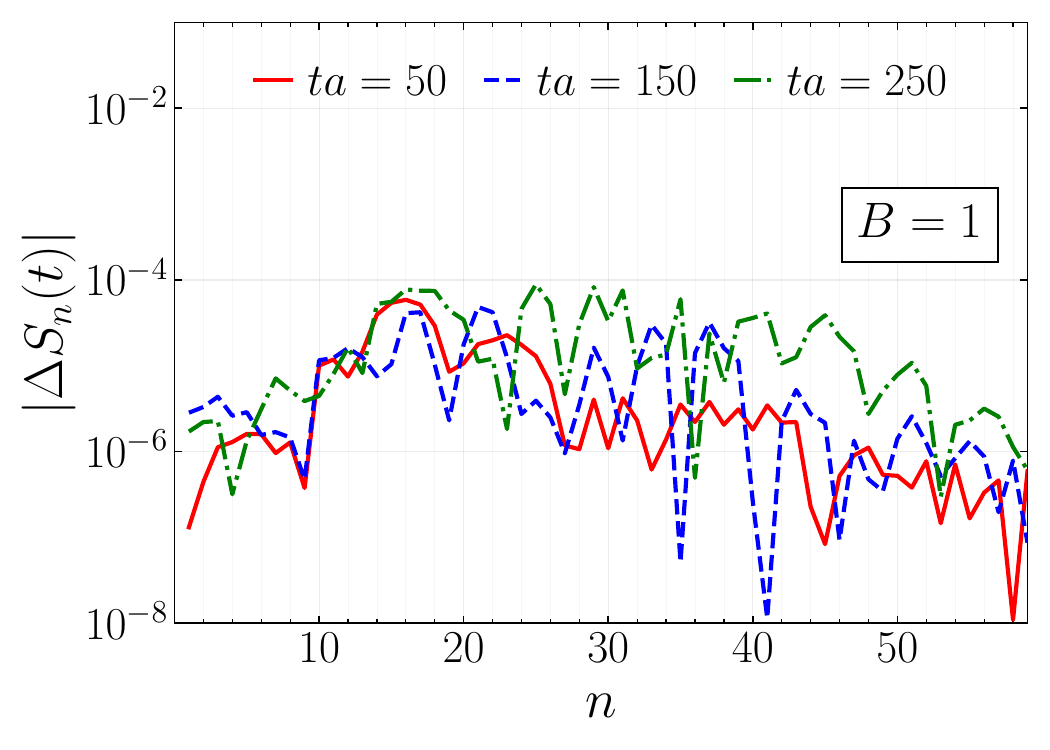}
    \includegraphics[width=0.45\textwidth]{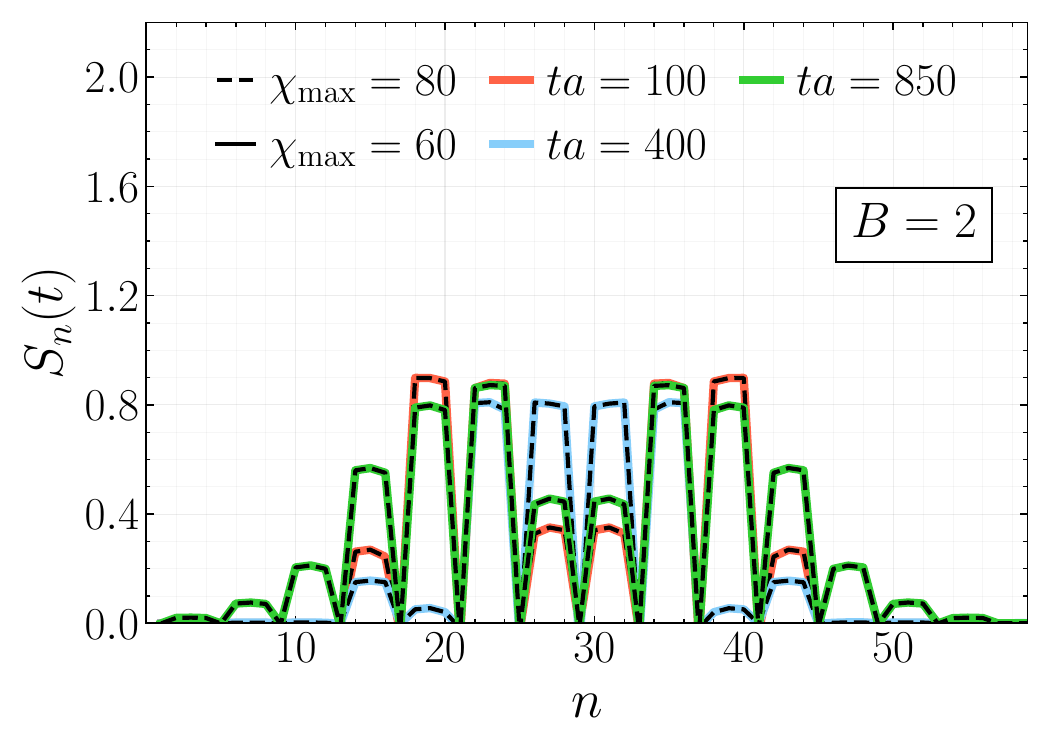}
    \includegraphics[width=0.45\textwidth]{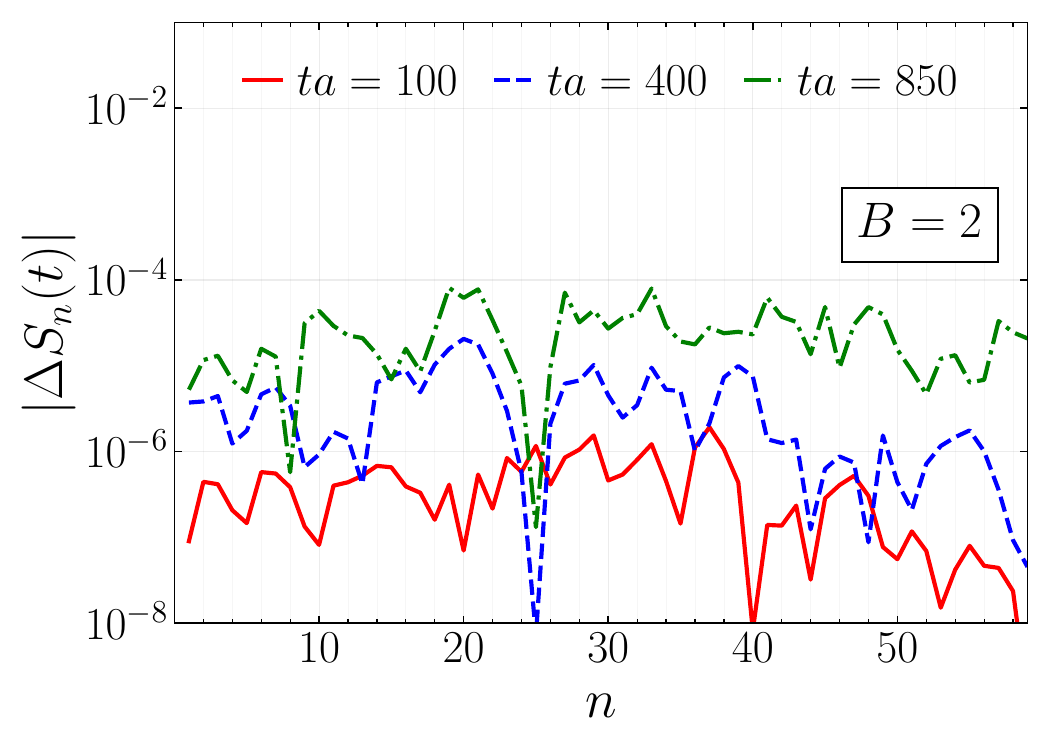}
    \caption{Two-site entanglement entropy $S_n(t)$ for $B=0$, $B=1$, and $B=2$ versus two different maximum bond dimension $\chi_\mathrm{max}=60$ and $80$ used in TDVP. In the right column, $|\Delta S_n(t)|=|S_n(t;\chi_\mathrm{max}=80)-S_n(t;\chi_\mathrm{max}=60)|$.
    }
\label{fig:error_bond_dimension}
\end{figure}

In Fig.~\ref{fig:error_bond_dimension}, we show the entanglement entropy $S_n(t)$ for different maximum bond dimensions $\chi_\mathrm{max}$ in the $B=0$, $B=1$, and $B=2$ scattering. We can see that $S_n(t)$ at selected time instances before, during, and after the scattering shows no qualitative dependences on the maximum bond dimensions. The associated error $|\Delta S_n(t)|=|S_n(t;\chi_\mathrm{max}=80)-S_n(t;\chi_\mathrm{max}=60)|$ calculated from $S_n(t)$ at two different $\chi_\mathrm{max}$ in the TDVP simulation is small, around $10^{-3}$, which is less than around 0.1\% compared to the entropy itself, though slightly increased error is observed for a longer simulation time. In summary, our choice of maximum bond dimension $\chi_\mathrm{max}=80$ used in the main result of the manuscript is sufficiently large enough and the same convergence pattern has been observed for other observables as well.  

\begin{figure}[h!]
    \centering
    \includegraphics[width=0.45\textwidth]{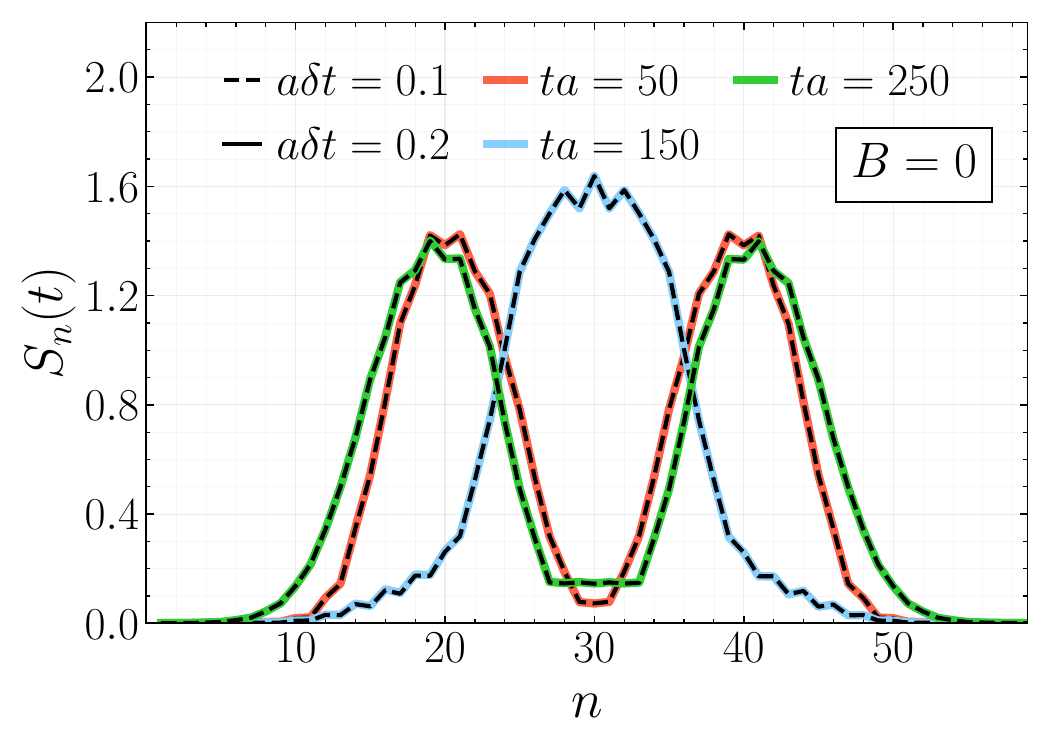}
    \includegraphics[width=0.45\textwidth]{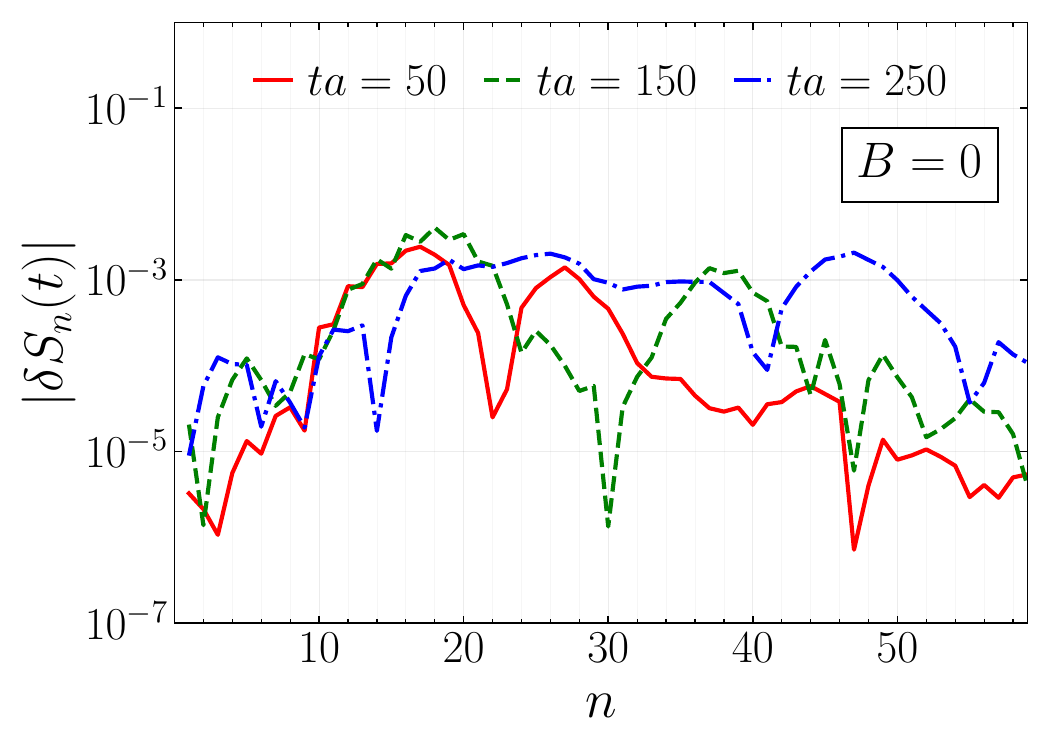}
    \includegraphics[width=0.45\textwidth]{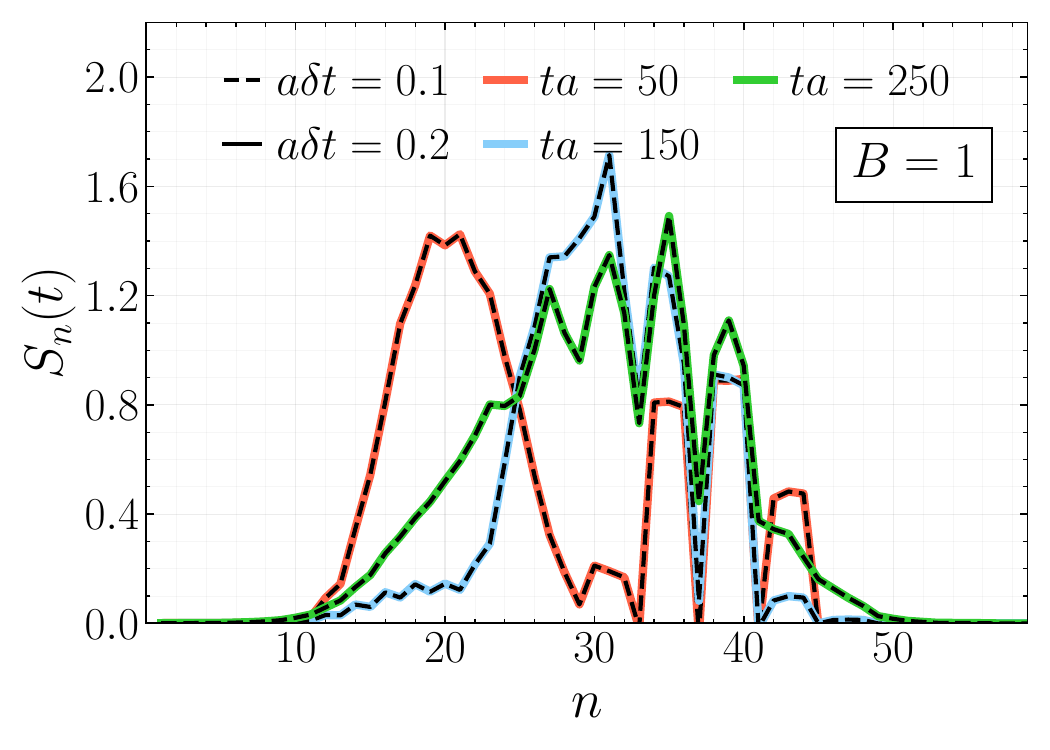}
    \includegraphics[width=0.45\textwidth]{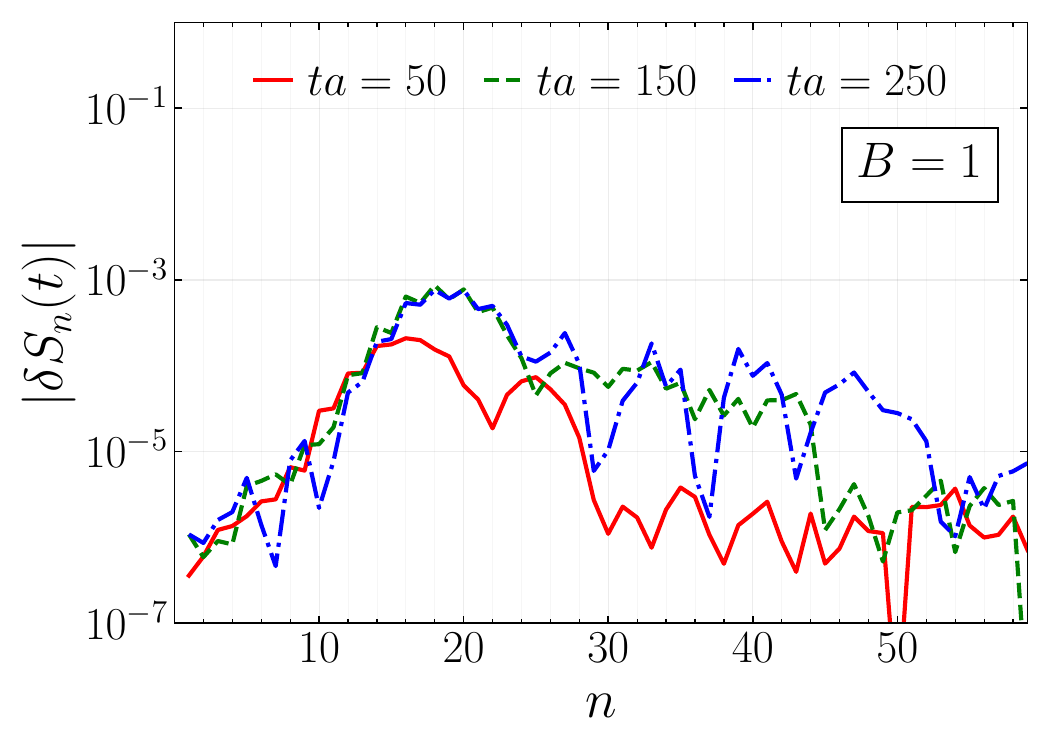}
    \includegraphics[width=0.45\textwidth]{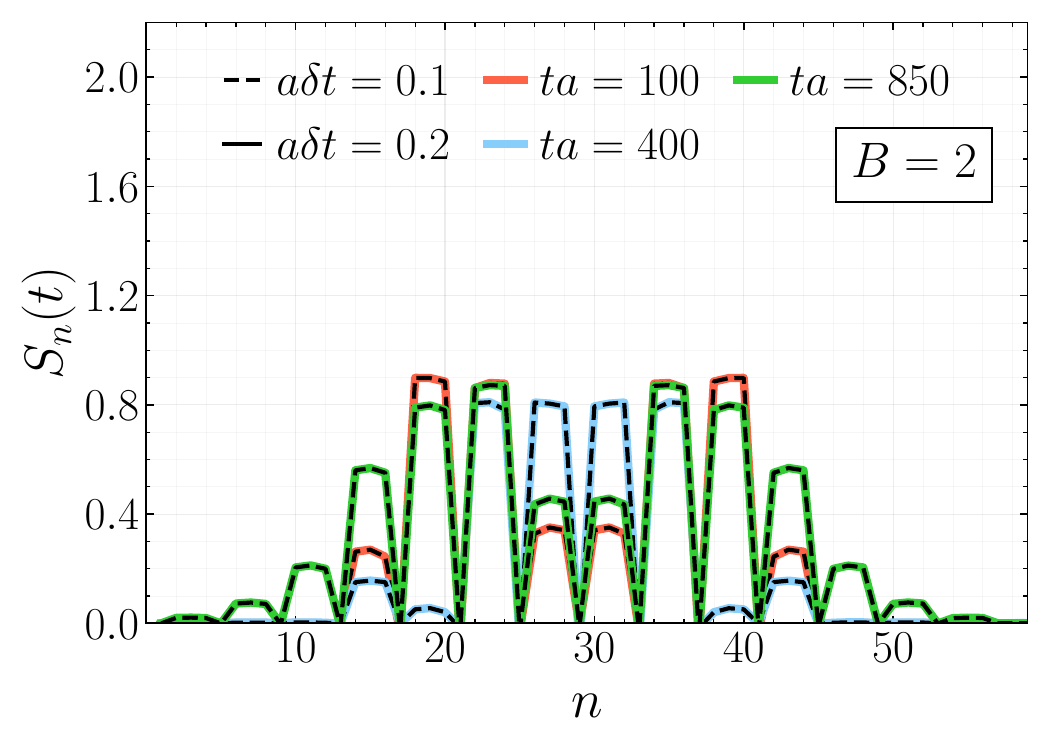}
    \includegraphics[width=0.45\textwidth]{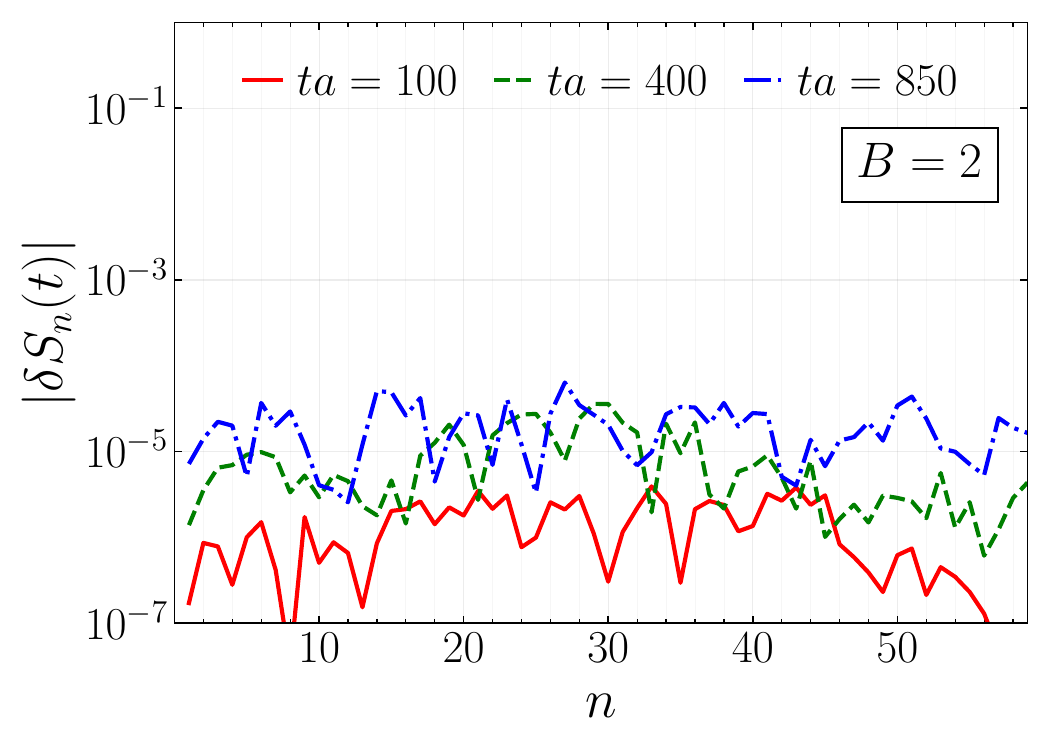}
    \caption{Two-site entanglement entropy $S_n(t)$ for $B=0$, $B=1$, and $B=2$ versus two different time step $a\delta t=0.1$ and $0.2$ used in TDVP. In the right column, $|\delta S_n(t)|=|S_n(t;a\delta t=0.1)-S_n(t;a\delta t=0.2)|$.
    }
\label{fig:error_dt}
\end{figure}

Next, we investigate our choice of time step used in the real-time evolution. In Fig.~\ref{fig:error_dt}, we show the entanglement entropy $S_n(t)$ for different time step $\a\delta t$ in the $B=0$, $B=1$, and $B=2$ scattering. Similarly, we see that $S_n(t)$ at selected time instances before, during, and after the scattering are not significantly affected by our choice of the time step, suggesting that $a\delta t=0.1$ is sufficiently small in our setup. We define the associated error $|\delta S_n(t)|=|S_n(t;a\delta t=0.1)-S_n(t;a\delta t=0.2)|$ to quantify the error from time step and $|\delta S_n(t)| \lesssim 10^{-3}$ is observed throughout the evolution. 

Last but not the least, it is also interesting and important to investigate the finite volume effects for the numerical simulation. The volume or system size of the setup is $\mathcal{V}=Nag$, where $N$ is the number of qubits, $a$ the lattice spacing, and $g$ the coupling strength. Currently our volume is fixed at $\mathcal{V}=300$ in the state preparation and in the real-time simulation. 
To truly investigate the finite volume effect on the real-time simulation as well as on the mass spectrum, one needs to increase $N$ and extrapolate the numerical results as $x\rightarrow \infty$, whose procedure is well demonstrated in Ref.~\cite{Banuls:2013jaa} for the U(1) Schwinger model. Similar, one also need to follow the same procedure to truly investigate finite volume effect for the real-time simulation. However, the SU(2) Hamiltonian employed in this work is much more difficult due to its non-Abelian nature, resulting in having more complicated internal structure and higher-order interactions. Rigorous study on the finite size effect goes beyond the scope of this work and probably is best suited to a different representation with gauge field truncation, and therefore we leave it for for the future. 

\section{Conclusion}\label{sec:conclusion}
We have presented a first study of real-time hadronic scattering in a $(1+1)$D SU(2) gauge theory using tensor network methods. 
We employed the lattice formulation of the theory in which the gauge degrees of freedom are explicitly integrated out, resulting in a long-range fermionic Hamiltonian that is free from any truncation of the electric field. 
Working close to the strong coupling limit, we analyzed scattering processes in different baryon number sectors, $B = 0, 1, 2$, observing that all dynamics are dominated by elastic channels. 
Nonetheless, for asymmetric kinematics we find that the slower of the two initial states becomes strongly delocalized after the collision, while the faster one propagates ballistically. 
This delocalization reflects the imbalance between kinetic and potential terms in the strong-coupling Hamiltonian and signals quasi-reflective interference rather than genuine inelasticity. 
The resulting correlation structure is well captured by the bipartite entanglement entropy and by the information lattice, which both show transient peaks during overlap followed by relaxation to their initial values. 

\begin{figure}
    \centering
\includegraphics[width=0.6\linewidth]{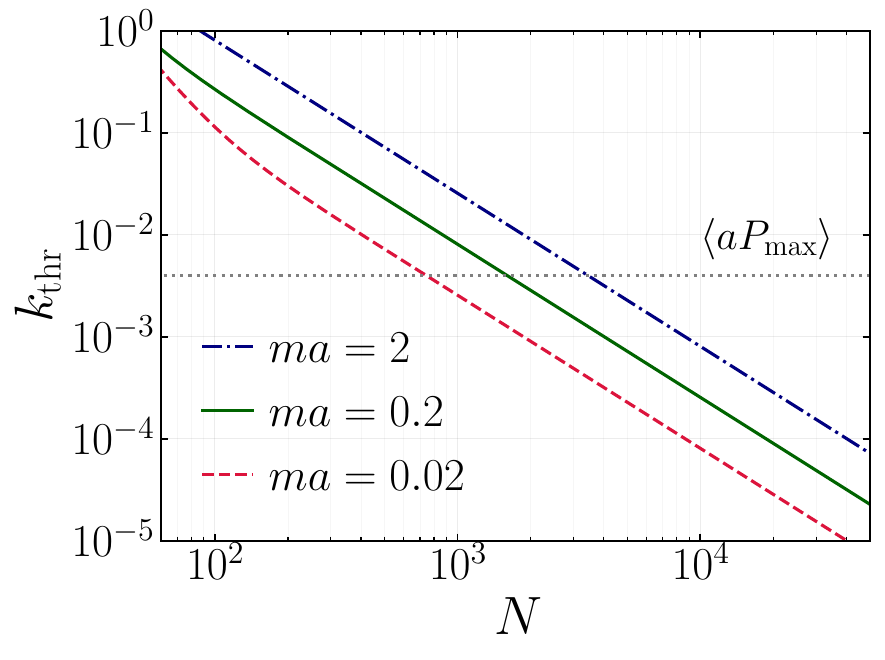}
    \caption{Threshold momentum of the baryon wavepacket as a function of the number of qubits $N$. The reference line of $\braket{aP_\mathrm{max}}=0.0048$ is provided for the baryon wavepacket of $ma=0.2$ considered in this work. 
    }
    \label{fig:scaling}
\end{figure}

Looking ahead, the present setup establishes the minimal requirements to study particle production and baryon-number \textit{rearrangement} from first principles in scattering processes. In this respect, one exciting prospect would be to study scattering of a baryon and an anti-baryon into two mesons ($M$), i.e., $B + \bar B \to M+M$, which, although conserving the total baryon number, results in the production of new final particle states. The minimal kinematic requirement for this transition is set by the threshold condition to produce the two final mesons: 
\begin{align}
p_{\rm thr}^2 = M_m^2 - M_b^2 \, , \quad
k_{\rm thr} = a p_{\rm thr} = \sqrt{(a M_m)^2 - (a M_b)^2} \, ,
\end{align}
which, using the strong-coupling estimates $M_b \simeq 2m$ and $M_m \simeq 2m + 3(ag)^2/8$, yields a dimensionless threshold lattice momentum 
\begin{align}
    k_{\rm thr}^2= 
    \left(2ma + \frac{3a}{8}(ag)^2\right)^2 - 4(ma)^2 \, .
\end{align}
Using the values quoted in the main text for $N=60$ qubits, we find that $k_{\rm thr} \simeq 0.66$, well above the maximum momentum for the baryon, $\braket{aP_{\mathrm{max}}} = 0.0048$, considered in this work. This is in agreement with the fact that we only observe elastic scattering. This threshold value can be lowered, for example, by increasing the number of qubits (at fixed volume) or by decreasing the fermion mass. We present representative estimates for this in Fig.~\ref{fig:scaling}. 

Although these parameters are beyond what can be simulated in the present implementation, several strategies can be pursued to reach the inelastic regime. Firstly, one can employ a formulation of the lattice theory in which local gauge degrees of freedom are explicitly retained, together with a mild truncation in the electric flux sector, see~\cite{Jakobs:2025rvz,Raychowdhury:2019iki,Grabowska:2024emw,DAndrea:2023qnr,Gustafson:2022xdt} for related discussion. Such a truncated representation makes it possible to explore regions of parameter space away from the deep strong-coupling limit, where the gap between meson and baryon states is smaller and inelastic channels become accessible at moderate lattice momenta. This extension will also require a modified strategy for the preparation of the initial states, for instance by constructing superpositions of localized baryon and anti-baryon wavepackets with tunable relative phases and momenta. Once inelastic processes are allowed, the model provides a natural framework to investigate how baryon number and color flux are dynamically redistributed during hadronic collisions. We leave the detailed study of these aspects to future work.

\section*{Acknowledgments}
We are grateful to Mari Carmen Bañuls, Meijian Li, and Enrique Rico for helpful discussions.
ZK is supported by National Science Foundation under grant No.~PHY-2515057. 
WQ is supported by the European Research Council under project ERC-2018-ADG-835105 YoctoLHC; by Maria de Maeztu excellence unit grant CEX2023-001318-M and project PID2020-119632GB-I00 funded by MICIU/AEI/10.13039/501100011033; by ERDF/EU; by the Marie Sklodowska-Curie Actions Fellowships under Grant No.~101109293; and by Xunta de Galicia (CIGUS Network of Research Centres).

\bibliographystyle{JHEP-2modlong.bst}
\bibliography{Lib.bib}

\end{document}